\documentclass[twocolumn,english,aps,prl,reprint,nofootinbib,superscriptaddress]{revtex4-2}
\pdfoutput=1
\usepackage[T1]{fontenc}
\usepackage[latin9]{inputenc}
\usepackage{amsmath}
\usepackage{amssymb}
\usepackage{graphicx}

\usepackage{color}
\usepackage{float}

\usepackage{tabularx}
\newcolumntype{Y}{>{\centering\arraybackslash}X}

\usepackage{multirow}

\usepackage{hyperref}

\hypersetup{
    colorlinks,
    linkcolor={black},
    citecolor={black},
    urlcolor={blue},
}

\makeatletter
\@ifundefined{textcolor}{}
{%
 \definecolor{BLACK}{gray}{0}
 \definecolor{WHITE}{gray}{1}
 \definecolor{RED}{rgb}{1,0,0}
 \definecolor{GREEN}{rgb}{0,1,0}
 \definecolor{BLUE}{rgb}{0,0,1}
 \definecolor{CYAN}{cmyk}{1,0,0,0}
 \definecolor{MAGENTA}{cmyk}{0,1,0,0}
 \definecolor{YELLOW}{cmyk}{0,0,1,0}
 \definecolor{PURPLE}{rgb}{0.7,0,0.7}
 \definecolor{dgreen}{rgb}{0,0.6,0}
}

\newcommand{\X}{X^{2}\Sigma^{+}}
\newcommand{\A}{A^{2}\Pi_{1/2}}
\newcommand{\B}{B^{2}\Sigma^{+}}
\newcommand{\wn}{~\rm{cm}^{-1}}

\usepackage{pslatex}

\usepackage{babel}

\makeatother

\begin{document}

\title{Vibronic branching ratios for nearly-closed rapid photon cycling of SrOH}

\author{Zack Lasner}
\email{zlasner@g.harvard.edu}
\affiliation{Harvard-MIT Center for Ultracold Atoms, Cambridge, MA 02138, USA}
\affiliation{Department of Physics, Harvard University, Cambridge, MA 02138, USA}

\author{Annika Lunstad}
\affiliation{Harvard-MIT Center for Ultracold Atoms, Cambridge, MA 02138, USA}
\affiliation{Department of Physics, Harvard University, Cambridge, MA 02138, USA}

\author{Chaoqun Zhang}
\affiliation{Department of Chemistry, The Johns Hopkins University, Baltimore, MD 21218, USA}

\author{Lan Cheng}
\affiliation{Department of Chemistry, The Johns Hopkins University, Baltimore, MD 21218, USA}

\author{John M. Doyle}
\email{jdoyle@g.harvard.edu}
\affiliation{Harvard-MIT Center for Ultracold Atoms, Cambridge, MA 02138, USA}
\affiliation{Department of Physics, Harvard University, Cambridge, MA 02138, USA}

\date{\today}






\begin{abstract}

The vibrational branching ratios of SrOH for radiative decay to the ground electronic state, $\X$, from the first two electronically excited states, $A^{2}\Pi$ and $\B$, are determined experimentally at the $\sim10^{-5}$ level.
The observed small branching ratios enable the design of a full, practical laser-cooling scheme, including magneto-optical trapping and sub-Doppler laser cooling, with $>10^4$ photon scatters per molecule. \emph{Ab initio} calculations sensitive to weak vibronic transitions are performed to facilitate the experimental measurement and analysis, and show good agreement with experiment.

\end{abstract}
\maketitle

\section{Introduction}

Experimentally observable time variation of fundamental constants are predicted by many dark matter models~\cite{arvanitaki2015searching,stadnik2015can,graham2013new,brdar2018fuzzy,banerjee2019coherent,stadnik2016improved,brzeminski2021time,cosme2018scale,marzola2018oscillating,krnjaic2018distorted}, inspiring diverse laboratory and cosmological searches~\cite{vanTilburg2015search,wcislo2016experimental,safronova2018two,hees2016searching,dzuba2021time,flambaum2006enhanced,zelevinsky2008precision,carollo2018two,geraci2016sensitivity,graham2016dark,badurina2022refined,arvanitaki2018search,pasteka2019material,branca2017search,manley2020searching,stadnik2015searching,geraci2019searching,kennedy2020precision,campbell2021searching,savalle2021searching,berlin2016neutrino,janish2020muon,vermeulen2021direct,hees2020search,choi2019new}. Recently, spectroscopy of SrOH vibrational states was proposed for measuring proton-to-electron mass ratio variation to probe dark matter candidate particles in the mass range of $10^{-22}-10^{-14}$~eV~\cite{kozyryev2021enhanced}, a region of notable theoretical and cosmological interest~\cite{hu2000fuzzy,marsh2015axion,lora2012mass}. Ultracold SrOH molecules would also be powerful probes for measurements of nuclear-spin-dependent parity violation~\cite{norrgard2019nuclear} and the electron electric dipole moment (eEDM), potentially improving upon current eEDM measurements by two orders of magnitude~\cite{kozyryev2017precision,gaul2020abinitio}.

All of the proposed experiments mentioned above rely on high precision spectroscopy, which greatly benefits from the low velocities and long interaction times provided by cooling molecules into the ultracold regime. SrOH was the first polyatomic molecule to be laser-cooled in one dimension~\cite{kozyryev2017sisyphus}, followed by CaOH~\cite{Baum2020magneto}, YbOH~\cite{augenbraun2020laser}, and CaOCH$_3$~\cite{Mitra2020direct}. Recently, a magneto-optical trap (MOT) was achieved for CaOH molecules~\cite{vilas2021magneto}, demonstrating a pathway that other polyatomic molecule laser cooling experiments can follow: molecules are produced in a cryogenic buffer gas beam, slowed via radiation pressure from counter-propagating laser light, and then captured into a MOT. These processes together generically require $\sim10^{4}$ photon scatters per molecule, necessitating the identification and repumping of vibrational states that are populated with probability $\gtrsim10^{-4}$ following laser excitation in the optical cycle~\cite{Baum2020establishing,zhang2021accurate}.

In this work, we present calculations and measurements of vibrational branching ratios (VBRs) of SrOH molecules upon excitation to the $\A(000)$, $\B(000)$, $\A(100)$, and $\B(100)$ states, where vibrational states are labeled $(v_{1}v_{2}^{\ell}v_{3})$, $v_1$ is the excitation of the Sr-O (symmetric) stretching mode, $v_{2}$ is the excitation of the Sr-O-H bending mode, $\ell\hbar$ is the angular momentum of the bending mode, and $v_{3}$ is the excitation of the O-H (anti-symmetric) stretching mode. We obtain sufficient experimental sensitivity to design a laser-cooling scheme with $>10^{4}$ photon scatters per molecule before leakage to an unaddressed vibrational state in the electronic ground state, $\X$. The calculations, which inform the target sensitivity of measurements and facilitate the estimation of branching ratios among spectroscopically unresolved decays, are in excellent agreement with experimental results. These measurements enable the first steps toward trapped, ultracold SrOH molecules.

\section{Theory}\label{sec:theory}

An optical cycle in an alkaline earth hydroxide such as SrOH involves transitions between the $\A$ or $\B$ states and 
the electronic ground $\X$ state. Spin-orbit coupling (SOC) and linear vibronic coupling (LVC) between the $\A$ and $\B$ states borrow intensities for nominally forbidden vibronic transitions to the $\X$ state, e.g., that from $\A(000)$ to $\X(010)$ \cite{Baum2020establishing,zhang2021accurate}. Although these transitions are relatively weak, at the level of $10^{-3}-10^{-2}$ or below, they play essential roles in laser-cooling experiments. To reliably guide such experiments, the computational treatment must therefore go beyond the Born-Oppenheimer approximation. Our calculations thus use a multi-state diabatic Hamiltonian \cite{Koeppel84} as constructed in Ref. \cite{zhang2021accurate} comprising the $\B$ state and the two degenerate $A^2\Pi$ states as well as the SOC and LVC among these states.

Previous calculations \cite{zhang2021accurate} show that SrOH features vibronic branching ratios smaller than those in CaOH for the transitions from the $\A(000)$ state to vibrational excited bending modes of the $\X$ state. This finding indicates that it is possible to form an optical cycle closed enough to enable loading of SrOH molecules into a MOT. Here we compute vibronic branching ratios for transitions from the $\B(000)$, $\B(100)$, and $\A(100)$ states to the vibrational levels of the $\X$ state to facilitate the design of an efficient repumping scheme. The discrete variable representation \cite{Colbert92} calculations for vibronic energies and wave functions employ the parameterization of the multi-state diabatic Hamiltonians as documented in Ref. \cite{zhang2021accurate}. This parameterization consists of the potential energy surfaces as well as SOC \cite{zhang2020performance} and LVC \cite{Ichino09} parameters computed at equation-of-motion coupled-cluster electron attachment singles and doubles level \cite{Stanton93a,Nooijen95} using triple-zeta basis sets \cite{Woon95,hill2017gaussian} with scalar-relativistic effects considered using spin-free exact two-component theory in its one-electron variant \cite{Dyall01,Liu2009,Cheng11b}. The present calculations apply a shift of 300 cm$^{-1}$ to the computed separation of the $\B$ and $A^2\Pi$ electronic states and obtain a separation of 1840 cm$^{-1}$ between the computed $\B(000)$ and $\A(000)$ levels, which compares favorably with the measured value of 1835 cm$^{-1}$~\cite{presunka1995laser,nakagawa1983high}. This shift helps capture the coupling of the $\B(000)$ state with the vibrational excited states of the $A^2\Pi$ state. We employ the CFOUR program package \cite{Matthews2020a,cfour} for all of the calculations presented here.

\section{Experiment}

\begin{figure}[t]
\begin{centering}
\includegraphics[width = 1\columnwidth]{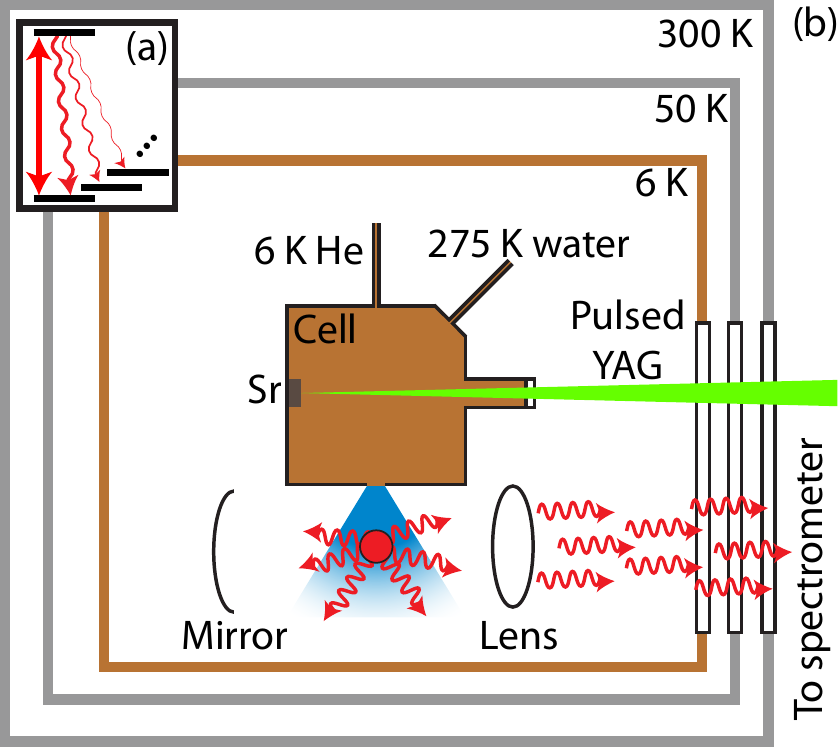}
\caption{\label{fig:apparatus} \textbf{(a)} Level diagram showing a single vibronic transition addressed by a laser (e.g., $X(000)\leftrightarrow A(000)$), resulting in decay to many vibrational states with sequentially smaller decay strengths. Stimulated absorption and emission from the laser is represented a straight double-headed arrow, while spontaneous emission is represented by wavy single-headed arrows. \textbf{(b)} Schematic of the apparatus viewed from above. A 6~K copper cell in a cryogenic chamber contains a strontium metal target, which is ablated by a pulsed Nd:YAG laser. The ablation products react with water to form SrOH, which is cooled by helium buffer gas and entrained in a beam out of the cell. A vertical retro-reflected laser excites molecules, which subsequently decay to numerous vibrational states. The fluorescence is collimated and directed toward the spectrometer outside of the cryogenic chamber.}
\par\end{centering}
\end{figure}

Designing a laser-cooling scheme sufficient for producing ultracold, trapped molecules necessitates identifying all decays whose probabilities sum to $\sim10^{-4}$. We therefore conservatively aim to determine all states populated by an optical cycling scheme with probability $\sim10^{-5}$ per spontaneous emission event. We measure VBRs from both $\A(000)$ and $\B(000)$ at the $\sim10^{-5}$ level of sensitivity in order to determine which state should be coupled to $\X(000)$ and hence contribute the majority of spontaneous emissions. In order to maximize the photon scattering rate, an optical cycle ideally does not couple excited vibrational states in $\X$ to the same excited state that is coupled to $\X(000)$. The dominant leakage channel--expected to be $\X(100)$ based on the calculations described in Sec.~\ref{sec:theory} and previous spectroscopy performed at the $\sim5\times10^{-3}$ level~\cite{nguyen2018fluorescence}--should therefore be coupled to whichever of $\A(000)$ and $\B(000)$ is not coupled to $\X(000)$ in the optical cycle. Finally, since calculations and previous spectroscopy suggest that fewer than 1\% of decays from $\A(000)$ or $\B(000)$ populate states other than $\X(000)$ or $\X(100)$, the VBRs from other excited states used for repumping can be determined at the level of only $\sim10^{-3}$. Therefore, VBRs from $\A(100)$ and $\B(100)$ are measured at the level of $\sim10^{-3}$ sensitivity.

The principle of the measurement is to excite molecules with a laser and compare the intensities of fluorescence to different vibrational states (see Fig.~\ref{fig:apparatus}(a)). Briefly, molecules in a cryogenic buffer gas beam are excited by a laser to the $J^{P}=1/2^{+}$ rotational state of $\A(000)$, $\B(000)$, $\A(100)$, or $\B(100)$, where $J$ is the total angular momentum excluding nuclear spin and $P$ is the parity. Fluorescence photons are dispersed on a Czerny-Turner style spectrometer and imaged on an electron-multiplying charge-coupled device (EMCCD). The intensities of observed wavelengths are proportional to the probability of populating corresponding vibrational states following a single photon scatter.

In more detail, cryogenic buffer gas beams of SrOH are produced in a copper cell held at $6$~K by a pulse tube refrigerator; see Fig.~\ref{fig:apparatus}(b). A pulsed Nd:YAG laser at 532 nm (20~mJ/pulse) ablates a strontium metal target to release atoms. Simultaneously, water vapor flows into the cell through a 275~K capillary and reacts with atoms to produce SrOH molecules. Cold helium thermalized with the cell walls cools the molecules and entrains them into a beam through a 7~mm diameter hole at the front of the cell.

At a distance of 2.5~cm from the cell exit, a retro-reflected laser beam propagating perpendicular to the direction of the molecular beam excites the $N^{P}=1^{-}$ state of $\X(000)$ to $E(v)$, where $N=J-S$ is the total angular momentum excluding spin, $E=\A$ or $\B$ is an electronic state, and $(v)=(000)$ or $(100)$. When the laser excites a state with $(v)=(000)$, the ground state spin-rotation splitting of 107 MHz is addressed with a sideband generated by an acousto-optic modulator (AOM). This closes the rotational structure of the optical cycle and allows up to $\sim$20 photons to be emitted per molecule (limited by decay to vibrationally excited states), proportionally increasing the fluorescence signal. When the laser excites a state with $(v)=(100)$, the molecules decay predominantly to $\X(100)$ after only a single photon scatter, and addressing both spin-rotation components of the ground $N^{P}=1^{-}$ state in $\X(000)$ would only increase the fluorescence signal by a factor of order unity. Therefore, we do not add a frequency sideband to the laser when exciting to $(v)=(100)$. The optical cycling (closed to the $\approx20$ photon level) when exciting to a $(v)=(000)$ state enables the target sensitivity of $\sim10^{-5}$ to be reached with fewer than 1\% as many experimental repetitions as would be required with no frequency sideband.

Molecular fluorescence is collimated by a $25$~mm diameter lens and sent to a 0.67~m Czerny-Turner style spectrometer. A 25~mm diameter mirror opposite the lens increases the collection efficiency. For each excited state, the observed emission wavelengths are linearly calibrated so that the peaks arising from decays to $\X(000)$ and $\X(200)$ agree with the known 1049~$\wn$ splitting of those states. The intensity sensitivity is calibrated as described in the SI of~\cite{zhang2021accurate} to an estimated 5\% relative uncertainty across wavelengths.

In order to isolate the signal arising from molecular fluorescence, it is necessary to subtract the backgrounds associated with scattered laser light, fluorescence from ambient excited-state strontium atoms, and EMCCD signal offsets. Therefore, the experiment is repeated for all combinations of the ablation laser ON or OFF, and the excitation light ON or OFF. Only the linear combination of images corresponding to molecular fluorescence is used to compute VBRs. The EMCCD can image approximately 39~nm of fluorescence dispersed from the grating at one time, so data taken over different wavelength regions are stitched together. Typically $\sim$15,000 experimental cycles are averaged for a particular wavelength range when exciting $(v)=(000)$, and $\sim$5,000 cycles are averaged when exciting $(v)=(100)$. Broad (tens of nm) drifts in the signal baseline at the level of $<10^{-4}$ decay probability are further fit and subtracted prior to calculation of VBRs.

\begin{figure}[t]
\begin{centering}
\includegraphics[width = 1\columnwidth]{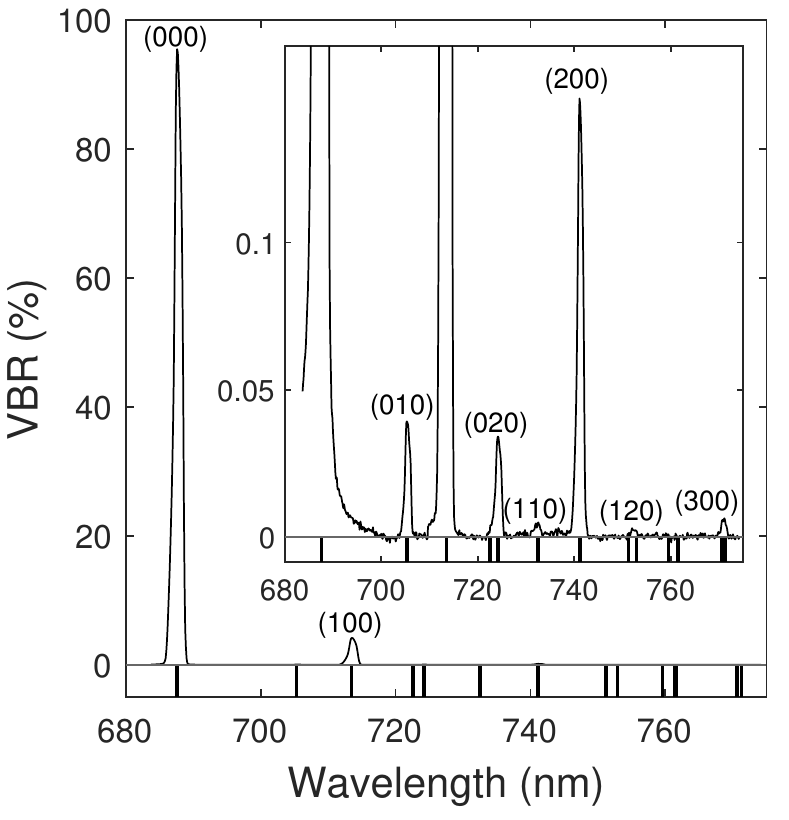}
\caption{\label{fig:A000-DLIF} Dispersed fluorescence spectrum upon laser excitation to $A(000)$, showing relative decay strengths to vibrational states in the $X$ electronic manifold. Labels above peaks identify the ground state vibrational level populated. Vertical bars below 0 denote the wavelengths of hypothetical emissions to all vibrational states of SrOH below 1630~$\wn$. Inset: Additional detail near the noise floor, showing decays with VBRs as small as 0.003\%.}
\par\end{centering}
\end{figure}

The dispersed fluorescence spectrum upon excitation to $\A(000)$ is shown in Fig.~\ref{fig:A000-DLIF}. Vibronic transitions are assigned based on the match to known or expected emission wavelengths. We assume the calculated ratios of transition probabilities among unresolved states (namely, emission to $(02^00)$ and $(02^20)$, $(12^00)$ and $(12^20)$, $(200)$ and $(03^10)$, or $(22^00)$ and $(05^10)$). Peak intensities are computed by integrating the signal over the width of a feature. Statistical uncertainties in the branching ratios are typically $\sim10^{-5}$ for $(v)=(000)$ states and $\sim 2\times10^{-3}$ for $(v)=(100)$ states, usually dominated by imperfections in background subtraction and EMCCD read noise. The uncertainty in a decay probability is computed by adding statistical uncertainty in quadrature with the systematic uncertainty of the system's wavelength-dependent intensity sensitivity. Calculated and measured VBRs from $E=\A$ are given in Table~\ref{tab:A-states}, and from $E=\B$ in Table~\ref{tab:B-states}. Vibrational state energies are assigned by fitting the peak locations relative to the $(000)$ peak, and are shown together with calculated energies in Table~\ref{tab:vibrational-energies}. Uncertainties in peak locations are obtained by adding wavelength calibration uncertainty in quadrature with statistical uncertainty in the fit location.

\section{Discussion}

\begin{table}
\begin{tabularx}{\columnwidth}{YYYY}
\toprule
  States & Calc. & High-resolution & This work \\
  \hline
  $(000)$ & 0 & 0 & 0 \\
  $(010)$ & 367 & 364~\cite{presunka1993high} & 364(3) \\
  $(100)$ & 534 & 527~\cite{presunka1995laser} & 528(2) \\
  $(02^{0}0)$ & 702 & 703~\cite{presunka1993high} & 699(4) \\
  $(02^{2}0)$ & 736 & 734~\cite{presunka1993high} & 735(4) \\
  $(110)$ & 892 & - & 883(7) \\
  $(03^{1}0)$ & 1048 & - & - \\
  $(200)$ & 1063 & 1049~\cite{presunka1995laser} & 1049 \\
  $(12^{0}0)$ & 1223 & - & 1215(9) \\
  $(12^{2}0)$ & 1254 & - & 1249(6) \\
  $(210)$ & 1411 & - & 1389(9) \\
  $(13^{1}0)$ & 1566 & - & - \\
  $(300)$ & 1595 & - & 1569(4) \\
  $(22^{0}0)$ & 1740 & - & - \\
  $(05^{1}0)$ & 1752 & - & 1720(8) \\
  $(15^{1}0)$ & 2269 & - & - \\
  \toprule
\end{tabularx}
\caption{\label{tab:vibrational-energies}Calculated and measured $\X$ vibrational energies in $\wn$. Energies known to $<1~\wn$ from high-resolution spectroscopy are also given where available. Experimental energies are linearly calibrated against known $(000)$ and $(200)$ energies.}
\end{table}

\begin{table}
\begin{tabularx}{\columnwidth}{YYYYY}
\toprule
  & \multicolumn{2}{c}{$A(000)$} & \multicolumn{2}{c}{$A(100)$}\\
  States & Calc. & Exp. & Calc. & Exp. \\
  \hline
  $\X(000)$ & 94.767 & 95.63(20) & 6.30 & 5.15(28) \\
  $\X(010)$ & 0.034 & 0.037(2) & - & - \\
  $\X(100)$ & 4.933 & 4.14(20) & 83.83 & 86.88(45) \\
  $\X(02^{0}0)$ & 0.012 & 0.008(1) & - & - \\
  $\X(02^{2}0)$ & 0.037 & 0.027(2) & - & - \\
  $\X(110)$ & 0.002 & 0.006(3) & - & - \\
  $\X(200)$ & 0.203 & 0.148(8) & 9.14 & 7.44(37) \\
  $\X(12^{2}0)$ & 0.003 & 0.003(1) & - & - \\
  $\X(300)$ & 0.008 & 0.006(1) & 0.57 & 0.54(15) \\
  \toprule
\end{tabularx}
\caption{\label{tab:A-states}Calculated and experimental VBRs for decay from $A$(000) and $A$(100). States are included if calculated or experimental values are at least $0.001\%$ or $0.1\%$ for $A(000)$ and $A(100)$, respectively.}
\end{table}

\begin{table}
\begin{tabularx}{\columnwidth}{YYYYY}
\toprule
  & \multicolumn{2}{c}{$B(000)$} & \multicolumn{2}{c}{$B(100)$}\\
  States & Calc. & Exp. & Calc. & Exp. \\
  \hline
  $\X(000)$ & 96.787 & 97.116(11) & 3.29 & 1.73(15) \\
  $\X(010)$ & 0.360 & 0.209(11) & - & - \\
  $\X(100)$ & 2.636 & 2.32(11) & 91.05 & 93.57(26) \\
  $\X(02^{0}0)$ & 0.037 & 0.125(6) & - & - \\
  $\X(110)$ & 0.010 & 0.039(2) & 0.34 & 0.20(5) \\
  $\X(03^{1}0)$ & 0.013 & 0.013(1) & - & - \\
  $\X(200)$ & 0.078 & 0.077(4) & 4.95 & 4.30(21) \\
  $\X(12^{0}0)$ & 0.008 & 0.015(2) & - & - \\
  $\X(210)$ & $<0.001$ & 0.006(2) & - & - \\
  $\X(13^{1}0)$ & 0.001 & 0.010(1) & - & - \\
  $\X(300)$ & 0.002 & 0.020(1) & 0.22 & 0.20(6) \\
  $\X(22^{0}0)$ & 0.001 & $<0.001$ & - & - \\
  $\X(05^{1}0)$ & 0.063 & 0.047(2) & - & - \\
  $\X(15^{1}0)$ & 0.003 & $<0.003$ & - & - \\
  \toprule
\end{tabularx}
\caption{\label{tab:B-states}Calculated and experimental VBRs for decay from $B$(000) and $B$(100). States are included if calculated or experimental values are at least $0.001\%$ or $0.1\%$ for $B(000)$ and $B(100)$, respectively. The assignment of $(210)$ is provisional.}
\end{table}

As seen in Table~\ref{tab:vibrational-energies}, the computed vibrational energies in $\X$ agree with experimental results to within 15~$\wn$ for states below $1400~\wn$ and to within $30~\wn$ for all states. The vibronic branching ratios reported in Tables~\ref{tab:A-states}-\ref{tab:B-states} in general show excellent agreement between experiment and calculation down to the level of $\sim10^{-5}$ branching probability. While the present calculations are helpful to guide the experimental design and enable assignments among spectroscopically unresolved decays, the highly sensitive measurements also provide guidance for future improvement of computations. The computed branching ratios decrease a bit too quickly with increasing vibrational quantum numbers for the $\B(000)\rightarrow\X$ decays. This results in underestimation for the intensities of the $\B(000)$ decays to $\X(02^00)$, $\X(110)$, and $\X(300)$ in Table \ref{tab:B-states}, as well as to a state at $1389(9)~\wn$. We tentatively assign this latter state to $(210)$, though it could plausibly be $(04^00)$ or $(04^20)$. It would be of interest to improve the computations by enhancing the quality of the potential energy surfaces with the inclusion of higher excitations in coupled-cluster calculations.

Furthermore, the calculations have difficulty capturing borrowing mechanisms due to near degeneracy of vibrational levels in electronic excited states. While the peak at $1569(4)~\wn$ in the measured $\A(000)\rightarrow\X$ spectrum originates exclusively from one transition that can be safely assigned to $\A(000)\rightarrow\X(300)$, the corresponding peak in the $\B(000)\rightarrow\X$ spectrum is atypically broad, with center located at $1554(10)~\wn$. We therefore speculate that the peak tentatively identified with $\B(000)\rightarrow\X(300)$ contains significant contribution from a second vibrational state $\sim20~\wn$ lower, presumably $\X(13^10)$, with a combined branching ratio of 0.030(2)\%. The nominally forbidden $\B(000)\rightarrow\X(13^10)$ transition borrows intensity from the coupling between the nearly degenerate $\B(000)$ and $A^2\Pi_{3/2}(13^10)$ states. Similarly, the $\B(000)\rightarrow\X(05^10)$ transition borrows intensity from the coupling between $\B(000)$ and $A^2\Pi_{1/2}(05^10)$ states. The computed intensities for these transitions are very sensitive to the relative level positions of vibrational states. The computations produce quite accurate intensity for the $\B(000)\rightarrow\X(05^10)$ transition as shown in Table \ref{tab:B-states}. However, they do not capture the coupling between $\B(000)$ and $A^2\Pi_{3/2}(13^10)$ accurately, and consequently underestimate the intensity for $\B(000)\rightarrow\X(13^10)$. Note that the computational combined intensity for $\B(000)$ decay to $\X(13^10)$ and $\X(300)$ is around one order of magnitude lower than the observed intensity.  Therefore, the calculations underestimate intensities for both of these spectroscopically unresolved transitions. Even if the 0.030\% VBR from $\B(000)$ to states around $1554~\wn$ could be attributed entirely to $\X(13^10)$, it would not limit the proposed photon cycling scheme described below. However, measurements of the relative contributions of $\B(000)\rightarrow\X(300)$ and $\B(000)\rightarrow\X(13^10)$ would still be of interest to provide benchmark results to guide refining calculations.

\begin{figure}[t]
\begin{centering}
\includegraphics[width = 1\columnwidth]{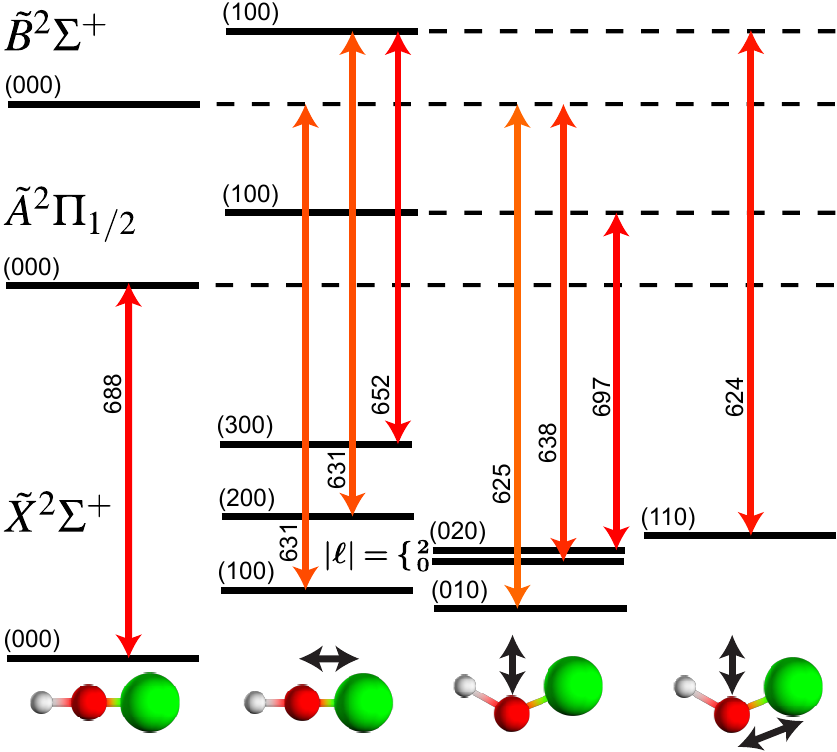}
\caption{\label{fig:cycling-scheme} Example laser-cooling scheme. On average,  more than 15,000 photon scatters occur before decay to an unaddressed vibrational state occurs. Vibrational states are organized by which modes are active; from left to right: none ($(v)=(000)$), Sr-O stretch ($v_{1}>0$), Sr-O-H bend ($v_{2}>0$), and hybrid stretch/bend ($v_{1},v_{2}>0$). Double-headed arrows denote the occurrence of both stimulated absorption and stimulated emission; hence, only one vibrational state is coupled to $\A(000)$ to achieve the maximum scattering rate. Laser wavelengths, in nm, are shown for each transition.}
\par\end{centering}
\end{figure}

The measured branching ratios are sufficient to design a laser cooling scheme that achieves more than $10^{4}$ photon scatters on average before a molecule decays to a vibrational state unaddressed by any lasers. Only one rotational state is populated for each vibrational state manifold except in $\X(010)$ and $\X(110)$, where both $N=1$ and $N=2$ states are populated (see~\cite{Baum2020establishing} for details). In those states, it is possible to repump with a frequency-modulated laser or to rapidly drive between states using microwave radiation while a repumping laser optically pumps one state. Alternatively, separate lasers might be used to address each rotational level. Therefore, the optical cycling scheme shown in Fig.~\ref{fig:cycling-scheme} may be implemented with 8--10 lasers, depending on what is most experimentally convenient. We also note that the proposed optical cycle maintains a high scattering rate by coupling only a single ground vibrational manifold to $\A(000)$, from which most decays occur.

To estimate the degree of closure of the optical cycle, we implement a Markov chain model similar to that described in~\cite{Baum2020establishing}. The transition probabilities from a vibrational state, $v$, to all other vibrational states are set equal to the VBRs from the excited state to which $v$ is coupled in Fig.~\ref{fig:cycling-scheme}. On average, a molecule is excited more than 15,000 times before decay to an unaddressed vibrational state, predominantly $(12^{2}0)$ and $(05^{1}0)$. The optical cycle is thus sufficiently closed to radiatively slow and magneto-optically trap molecules following an experimental protocol like that used for CaOH. Remarkably, we find that this optical cycle in SrOH achieves a higher number of scattered photons than that used to trap CaOH, while requiring fewer lasers~\cite{vilas2021magneto}.

We emphasize that, as has been found with CaOH and YbOH~\cite{zhang2021accurate}, extrapolating vibrational branching trends from the first few populated vibrational states does not accurately predict which states are populated after $\sim10^{4}$ photon scatters, due to small perturbations in the molecular Hamiltonian described in Sec.~\ref{sec:theory}. Thus, while the seminal laser cooling work in polyatomic molecules achieved one-dimensional Sisyphus cooling of SrOH using only 200 photons per molecule, our results provide the first definitive pathway to an optical cycle that is sufficiently closed to produce ultracold, trapped SrOH molecules.

In summary, we have measured vibrational branching ratios at $\sim10^{-5}$ relative sensitivity in SrOH, for both the $\A(000)$ and $\B(000)$ excited states. Additional measurements at $\sim10^{-3}$ relative sensitivity for $\A(100)$ and $\B(100)$ enable a full optical cycling scheme to be designed for more than 15,000 photon scatters per molecule. We have also performed high-accuracy calculations of vibronic structure and vibrational branching ratios, which are validated by good agreement with measurements and are used to assign vibrational branching ratios among unresolved fluorescence peaks. This work is a critical step to implementing an optical cycle capable of producing trapped, ultracold SrOH molecules for measuring proton-to-electron mass ratio variation.

\begin{acknowledgments}

We are grateful to Christian Hallas and Nathaniel Vilas for sharing critical experimental equipment, as well as to Benjamin Augenbraun and Yicheng Bao for helpful discussions. We thank Ivan Kozyryev for valuable feedback on the manuscript. This work was supported by an NSF QLCI award as part of the Q-SEnSE institute. The computational work at Johns Hopkins University was supported by National Science Foundation under Grant No. PHY-2011794.

\end{acknowledgments}

\bibliographystyle{apsrev4-2}
\bibliography{SrOH_VBRs}

\begin{thebibliography}{66}%
\makeatletter
\providecommand \@ifxundefined [1]{%
 \@ifx{#1\undefined}
}%
\providecommand \@ifnum [1]{%
 \ifnum #1\expandafter \@firstoftwo
 \else \expandafter \@secondoftwo
 \fi
}%
\providecommand \@ifx [1]{%
 \ifx #1\expandafter \@firstoftwo
 \else \expandafter \@secondoftwo
 \fi
}%
\providecommand \natexlab [1]{#1}%
\providecommand \enquote  [1]{``#1''}%
\providecommand \bibnamefont  [1]{#1}%
\providecommand \bibfnamefont [1]{#1}%
\providecommand \citenamefont [1]{#1}%
\providecommand \href@noop [0]{\@secondoftwo}%
\providecommand \href [0]{\begingroup \@sanitize@url \@href}%
\providecommand \@href[1]{\@@startlink{#1}\@@href}%
\providecommand \@@href[1]{\endgroup#1\@@endlink}%
\providecommand \@sanitize@url [0]{\catcode `\\12\catcode `\$12\catcode
  `\&12\catcode `\#12\catcode `\^12\catcode `\_12\catcode `\%12\relax}%
\providecommand \@@startlink[1]{}%
\providecommand \@@endlink[0]{}%
\providecommand \url  [0]{\begingroup\@sanitize@url \@url }%
\providecommand \@url [1]{\endgroup\@href {#1}{\urlprefix }}%
\providecommand \urlprefix  [0]{URL }%
\providecommand \Eprint [0]{\href }%
\providecommand \doibase [0]{https://doi.org/}%
\providecommand \selectlanguage [0]{\@gobble}%
\providecommand \bibinfo  [0]{\@secondoftwo}%
\providecommand \bibfield  [0]{\@secondoftwo}%
\providecommand \translation [1]{[#1]}%
\providecommand \BibitemOpen [0]{}%
\providecommand \bibitemStop [0]{}%
\providecommand \bibitemNoStop [0]{.\EOS\space}%
\providecommand \EOS [0]{\spacefactor3000\relax}%
\providecommand \BibitemShut  [1]{\csname bibitem#1\endcsname}%
\let\auto@bib@innerbib\@empty
\bibitem [{\citenamefont {Arvanitaki}\ \emph {et~al.}(2015)\citenamefont
  {Arvanitaki}, \citenamefont {Huang},\ and\ \citenamefont
  {Van~Tilburg}}]{arvanitaki2015searching}%
  \BibitemOpen
  \bibfield  {author} {\bibinfo {author} {\bibfnamefont {A.}~\bibnamefont
  {Arvanitaki}}, \bibinfo {author} {\bibfnamefont {J.}~\bibnamefont {Huang}},\
  and\ \bibinfo {author} {\bibfnamefont {K.}~\bibnamefont {Van~Tilburg}},\
  }\href {https://doi.org/10.1103/PhysRevD.91.015015} {\bibfield  {journal}
  {\bibinfo  {journal} {Phys. Rev. D}\ }\textbf {\bibinfo {volume} {91}},\
  \bibinfo {pages} {015015} (\bibinfo {year} {2015})}\BibitemShut {NoStop}%
\bibitem [{\citenamefont {Stadnik}\ and\ \citenamefont
  {Flambaum}(2015{\natexlab{a}})}]{stadnik2015can}%
  \BibitemOpen
  \bibfield  {author} {\bibinfo {author} {\bibfnamefont {Y.}~\bibnamefont
  {Stadnik}}\ and\ \bibinfo {author} {\bibfnamefont {V.}~\bibnamefont
  {Flambaum}},\ }\href {https://doi.org/10.1103/PhysRevLett.115.201301}
  {\bibfield  {journal} {\bibinfo  {journal} {Phys. Rev. Lett.}\ }\textbf
  {\bibinfo {volume} {115}},\ \bibinfo {pages} {201301} (\bibinfo {year}
  {2015}{\natexlab{a}})}\BibitemShut {NoStop}%
\bibitem [{\citenamefont {Graham}\ and\ \citenamefont
  {Rajendran}(2013)}]{graham2013new}%
  \BibitemOpen
  \bibfield  {author} {\bibinfo {author} {\bibfnamefont {P.~W.}\ \bibnamefont
  {Graham}}\ and\ \bibinfo {author} {\bibfnamefont {S.}~\bibnamefont
  {Rajendran}},\ }\href {https://doi.org/10.1103/PhysRevD.88.035023} {\bibfield
   {journal} {\bibinfo  {journal} {Phys. Rev. D}\ }\textbf {\bibinfo {volume}
  {88}},\ \bibinfo {pages} {035023} (\bibinfo {year} {2013})}\BibitemShut
  {NoStop}%
\bibitem [{\citenamefont {Brdar}\ \emph {et~al.}(2018)\citenamefont {Brdar},
  \citenamefont {Kopp}, \citenamefont {Liu}, \citenamefont {Prass},\ and\
  \citenamefont {Wang}}]{brdar2018fuzzy}%
  \BibitemOpen
  \bibfield  {author} {\bibinfo {author} {\bibfnamefont {V.}~\bibnamefont
  {Brdar}}, \bibinfo {author} {\bibfnamefont {J.}~\bibnamefont {Kopp}},
  \bibinfo {author} {\bibfnamefont {J.}~\bibnamefont {Liu}}, \bibinfo {author}
  {\bibfnamefont {P.}~\bibnamefont {Prass}},\ and\ \bibinfo {author}
  {\bibfnamefont {X.~P.}\ \bibnamefont {Wang}},\ }\href
  {https://doi.org/10.1103/PhysRevD.97.043001} {\bibfield  {journal} {\bibinfo
  {journal} {Phys. Rev. D}\ }\textbf {\bibinfo {volume} {97}},\ \bibinfo
  {pages} {43001} (\bibinfo {year} {2018})}\BibitemShut {NoStop}%
\bibitem [{\citenamefont {Banerjee}\ \emph {et~al.}(2019)\citenamefont
  {Banerjee}, \citenamefont {Kim},\ and\ \citenamefont
  {Perez}}]{banerjee2019coherent}%
  \BibitemOpen
  \bibfield  {author} {\bibinfo {author} {\bibfnamefont {A.}~\bibnamefont
  {Banerjee}}, \bibinfo {author} {\bibfnamefont {H.}~\bibnamefont {Kim}},\ and\
  \bibinfo {author} {\bibfnamefont {G.}~\bibnamefont {Perez}},\ }\href
  {https://doi.org/10.1103/PhysRevD.100.115026} {\bibfield  {journal} {\bibinfo
   {journal} {Phys. Rev. D}\ }\textbf {\bibinfo {volume} {100}},\ \bibinfo
  {pages} {115026} (\bibinfo {year} {2019})}\BibitemShut {NoStop}%
\bibitem [{\citenamefont {Stadnik}\ and\ \citenamefont
  {Flambaum}(2016)}]{stadnik2016improved}%
  \BibitemOpen
  \bibfield  {author} {\bibinfo {author} {\bibfnamefont {Y.~V.}\ \bibnamefont
  {Stadnik}}\ and\ \bibinfo {author} {\bibfnamefont {V.~V.}\ \bibnamefont
  {Flambaum}},\ }\href {https://doi.org/10.1103/PhysRevA.94.022111} {\bibfield
  {journal} {\bibinfo  {journal} {Phys. Rev. A}\ }\textbf {\bibinfo {volume}
  {94}},\ \bibinfo {pages} {022111} (\bibinfo {year} {2016})}\BibitemShut
  {NoStop}%
\bibitem [{\citenamefont {Brzeminski}\ \emph {et~al.}(2021)\citenamefont
  {Brzeminski}, \citenamefont {Chacko}, \citenamefont {Dev},\ and\
  \citenamefont {Hook}}]{brzeminski2021time}%
  \BibitemOpen
  \bibfield  {author} {\bibinfo {author} {\bibfnamefont {D.}~\bibnamefont
  {Brzeminski}}, \bibinfo {author} {\bibfnamefont {Z.}~\bibnamefont {Chacko}},
  \bibinfo {author} {\bibfnamefont {A.}~\bibnamefont {Dev}},\ and\ \bibinfo
  {author} {\bibfnamefont {A.}~\bibnamefont {Hook}},\ }\href
  {https://doi.org/10.1103/PhysRevD.104.075019} {\bibfield  {journal} {\bibinfo
   {journal} {Phys. Rev. D}\ }\textbf {\bibinfo {volume} {104}},\ \bibinfo
  {pages} {75019} (\bibinfo {year} {2021})}\BibitemShut {NoStop}%
\bibitem [{\citenamefont {Cosme}\ \emph {et~al.}(2018)\citenamefont {Cosme},
  \citenamefont {Rosa},\ and\ \citenamefont {Bertolami}}]{cosme2018scale}%
  \BibitemOpen
  \bibfield  {author} {\bibinfo {author} {\bibfnamefont {C.}~\bibnamefont
  {Cosme}}, \bibinfo {author} {\bibfnamefont {J.~G.}\ \bibnamefont {Rosa}},\
  and\ \bibinfo {author} {\bibfnamefont {O.}~\bibnamefont {Bertolami}},\ }\href
  {https://doi.org/10.1007/JHEP05(2018)129} {\bibfield  {journal} {\bibinfo
  {journal} {J. High Energ. Phys.}\ }\textbf {\bibinfo {volume}
  {2018}}}\BibitemShut {NoStop}%
\bibitem [{\citenamefont {Marzola}\ \emph {et~al.}(2018)\citenamefont
  {Marzola}, \citenamefont {Raidal},\ and\ \citenamefont
  {Urban}}]{marzola2018oscillating}%
  \BibitemOpen
  \bibfield  {author} {\bibinfo {author} {\bibfnamefont {L.}~\bibnamefont
  {Marzola}}, \bibinfo {author} {\bibfnamefont {M.}~\bibnamefont {Raidal}},\
  and\ \bibinfo {author} {\bibfnamefont {F.~R.}\ \bibnamefont {Urban}},\ }\href
  {https://doi.org/10.1103/PhysRevD.97.024010} {\bibfield  {journal} {\bibinfo
  {journal} {Phys. Rev. D}\ }\textbf {\bibinfo {volume} {97}},\ \bibinfo
  {pages} {24010} (\bibinfo {year} {2018})}\BibitemShut {NoStop}%
\bibitem [{\citenamefont {Krnjaic}\ \emph {et~al.}(2018)\citenamefont
  {Krnjaic}, \citenamefont {Machado},\ and\ \citenamefont
  {Necib}}]{krnjaic2018distorted}%
  \BibitemOpen
  \bibfield  {author} {\bibinfo {author} {\bibfnamefont {G.}~\bibnamefont
  {Krnjaic}}, \bibinfo {author} {\bibfnamefont {P.~A.}\ \bibnamefont
  {Machado}},\ and\ \bibinfo {author} {\bibfnamefont {L.}~\bibnamefont
  {Necib}},\ }\href {https://doi.org/10.1103/PhysRevD.97.075017} {\bibfield
  {journal} {\bibinfo  {journal} {Phys. Rev. D}\ }\textbf {\bibinfo {volume}
  {97}},\ \bibinfo {pages} {75017} (\bibinfo {year} {2018})}\BibitemShut
  {NoStop}%
\bibitem [{\citenamefont {{Van Tilburg}}\ \emph {et~al.}(2015)\citenamefont
  {{Van Tilburg}}, \citenamefont {Leefer}, \citenamefont {Bougas},\ and\
  \citenamefont {Budker}}]{vanTilburg2015search}%
  \BibitemOpen
  \bibfield  {author} {\bibinfo {author} {\bibfnamefont {K.}~\bibnamefont {{Van
  Tilburg}}}, \bibinfo {author} {\bibfnamefont {N.}~\bibnamefont {Leefer}},
  \bibinfo {author} {\bibfnamefont {L.}~\bibnamefont {Bougas}},\ and\ \bibinfo
  {author} {\bibfnamefont {D.}~\bibnamefont {Budker}},\ }\href
  {https://doi.org/10.1103/PHYSREVLETT.115.011802} {\bibfield  {journal}
  {\bibinfo  {journal} {Phys. Rev. Lett.}\ }\textbf {\bibinfo {volume} {115}},\
  \bibinfo {pages} {011802} (\bibinfo {year} {2015})}\BibitemShut {NoStop}%
\bibitem [{\citenamefont {Wcislo}\ \emph {et~al.}(2016)\citenamefont {Wcislo},
  \citenamefont {Morzynski}, \citenamefont {Bober}, \citenamefont {Cygan},
  \citenamefont {Lisak}, \citenamefont {Ciurylo},\ and\ \citenamefont
  {Zawada}}]{wcislo2016experimental}%
  \BibitemOpen
  \bibfield  {author} {\bibinfo {author} {\bibfnamefont {P.}~\bibnamefont
  {Wcislo}}, \bibinfo {author} {\bibfnamefont {P.}~\bibnamefont {Morzynski}},
  \bibinfo {author} {\bibfnamefont {M.}~\bibnamefont {Bober}}, \bibinfo
  {author} {\bibfnamefont {A.}~\bibnamefont {Cygan}}, \bibinfo {author}
  {\bibfnamefont {D.}~\bibnamefont {Lisak}}, \bibinfo {author} {\bibfnamefont
  {R.}~\bibnamefont {Ciurylo}},\ and\ \bibinfo {author} {\bibfnamefont
  {M.}~\bibnamefont {Zawada}},\ }\href
  {https://doi.org/10.1038/s41550-016-0009} {\bibfield  {journal} {\bibinfo
  {journal} {Nat. Astron.}\ }\textbf {\bibinfo {volume} {1}},\ \bibinfo {pages}
  {0009} (\bibinfo {year} {2016})}\BibitemShut {NoStop}%
\bibitem [{\citenamefont {Safronova}\ \emph {et~al.}(2018)\citenamefont
  {Safronova}, \citenamefont {Porsev}, \citenamefont {Sanner},\ and\
  \citenamefont {Ye}}]{safronova2018two}%
  \BibitemOpen
  \bibfield  {author} {\bibinfo {author} {\bibfnamefont {M.~S.}\ \bibnamefont
  {Safronova}}, \bibinfo {author} {\bibfnamefont {S.~G.}\ \bibnamefont
  {Porsev}}, \bibinfo {author} {\bibfnamefont {C.}~\bibnamefont {Sanner}},\
  and\ \bibinfo {author} {\bibfnamefont {J.}~\bibnamefont {Ye}},\ }\href
  {https://doi.org/10.1103/PhysRevLett.120.173001} {\bibfield  {journal}
  {\bibinfo  {journal} {Phys. Rev. Lett.}\ }\textbf {\bibinfo {volume} {120}},\
  \bibinfo {pages} {173001} (\bibinfo {year} {2018})}\BibitemShut {NoStop}%
\bibitem [{\citenamefont {Hees}\ \emph {et~al.}(2016)\citenamefont {Hees},
  \citenamefont {Gu{\'e}na}, \citenamefont {Abgrall}, \citenamefont {Bize},\
  and\ \citenamefont {Wolf}}]{hees2016searching}%
  \BibitemOpen
  \bibfield  {author} {\bibinfo {author} {\bibfnamefont {A.}~\bibnamefont
  {Hees}}, \bibinfo {author} {\bibfnamefont {J.}~\bibnamefont {Gu{\'e}na}},
  \bibinfo {author} {\bibfnamefont {M.}~\bibnamefont {Abgrall}}, \bibinfo
  {author} {\bibfnamefont {S.}~\bibnamefont {Bize}},\ and\ \bibinfo {author}
  {\bibfnamefont {P.}~\bibnamefont {Wolf}},\ }\href
  {https://doi.org/10.1103/PhysRevLett.117.061301} {\bibfield  {journal}
  {\bibinfo  {journal} {Phys. Rev. Lett.}\ }\textbf {\bibinfo {volume} {117}},\
  \bibinfo {pages} {061301} (\bibinfo {year} {2016})}\BibitemShut {NoStop}%
\bibitem [{\citenamefont {Dzuba}\ \emph {et~al.}(2021)\citenamefont {Dzuba},
  \citenamefont {Allehabi}, \citenamefont {Flambaum}, \citenamefont {Li},\ and\
  \citenamefont {Schiller}}]{dzuba2021time}%
  \BibitemOpen
  \bibfield  {author} {\bibinfo {author} {\bibfnamefont {V.~A.}\ \bibnamefont
  {Dzuba}}, \bibinfo {author} {\bibfnamefont {S.~O.}\ \bibnamefont {Allehabi}},
  \bibinfo {author} {\bibfnamefont {V.~V.}\ \bibnamefont {Flambaum}}, \bibinfo
  {author} {\bibfnamefont {J.}~\bibnamefont {Li}},\ and\ \bibinfo {author}
  {\bibfnamefont {S.}~\bibnamefont {Schiller}},\ }\href
  {https://doi.org/10.1103/PHYSREVA.103.022822} {\bibfield  {journal} {\bibinfo
   {journal} {Phys. Rev. A}\ }\textbf {\bibinfo {volume} {103}},\ \bibinfo
  {pages} {022822} (\bibinfo {year} {2021})}\BibitemShut {NoStop}%
\bibitem [{\citenamefont {Flambaum}(2006)}]{flambaum2006enhanced}%
  \BibitemOpen
  \bibfield  {author} {\bibinfo {author} {\bibfnamefont {V.~V.}\ \bibnamefont
  {Flambaum}},\ }\href {https://doi.org/10.1103/PHYSREVLETT.97.092502}
  {\bibfield  {journal} {\bibinfo  {journal} {Phys. Rev. Lett.}\ }\textbf
  {\bibinfo {volume} {97}},\ \bibinfo {pages} {092502} (\bibinfo {year}
  {2006})}\BibitemShut {NoStop}%
\bibitem [{\citenamefont {Zelevinsky}\ \emph {et~al.}(2008)\citenamefont
  {Zelevinsky}, \citenamefont {Kotochigova},\ and\ \citenamefont
  {Ye}}]{zelevinsky2008precision}%
  \BibitemOpen
  \bibfield  {author} {\bibinfo {author} {\bibfnamefont {T.}~\bibnamefont
  {Zelevinsky}}, \bibinfo {author} {\bibfnamefont {S.}~\bibnamefont
  {Kotochigova}},\ and\ \bibinfo {author} {\bibfnamefont {J.}~\bibnamefont
  {Ye}},\ }\href {https://doi.org/10.1103/PhysRevLett.100.043201} {\bibfield
  {journal} {\bibinfo  {journal} {Phys. Rev. Lett.}\ }\textbf {\bibinfo
  {volume} {100}},\ \bibinfo {pages} {043201} (\bibinfo {year}
  {2008})}\BibitemShut {NoStop}%
\bibitem [{\citenamefont {Carollo}\ \emph {et~al.}(2018)\citenamefont
  {Carollo}, \citenamefont {Frenett},\ and\ \citenamefont
  {Hanneke}}]{carollo2018two}%
  \BibitemOpen
  \bibfield  {author} {\bibinfo {author} {\bibfnamefont {R.}~\bibnamefont
  {Carollo}}, \bibinfo {author} {\bibfnamefont {A.}~\bibnamefont {Frenett}},\
  and\ \bibinfo {author} {\bibfnamefont {D.}~\bibnamefont {Hanneke}},\ }\href
  {https://doi.org/10.3390/atoms7010001} {\bibfield  {journal} {\bibinfo
  {journal} {Atoms}\ }\textbf {\bibinfo {volume} {7}},\ \bibinfo {pages} {1}
  (\bibinfo {year} {2018})}\BibitemShut {NoStop}%
\bibitem [{\citenamefont {Geraci}\ and\ \citenamefont
  {Derevianko}(2016)}]{geraci2016sensitivity}%
  \BibitemOpen
  \bibfield  {author} {\bibinfo {author} {\bibfnamefont {A.~A.}\ \bibnamefont
  {Geraci}}\ and\ \bibinfo {author} {\bibfnamefont {A.}~\bibnamefont
  {Derevianko}},\ }\href {https://doi.org/10.1103/PhysRevLett.117.261301}
  {\bibfield  {journal} {\bibinfo  {journal} {Phys. Rev. Lett.}\ }\textbf
  {\bibinfo {volume} {117}},\ \bibinfo {pages} {261301} (\bibinfo {year}
  {2016})}\BibitemShut {NoStop}%
\bibitem [{\citenamefont {Graham}\ \emph {et~al.}(2016)\citenamefont {Graham},
  \citenamefont {Kaplan}, \citenamefont {Mardon}, \citenamefont {Rajendran},\
  and\ \citenamefont {Terrano}}]{graham2016dark}%
  \BibitemOpen
  \bibfield  {author} {\bibinfo {author} {\bibfnamefont {P.~W.}\ \bibnamefont
  {Graham}}, \bibinfo {author} {\bibfnamefont {D.~E.}\ \bibnamefont {Kaplan}},
  \bibinfo {author} {\bibfnamefont {J.}~\bibnamefont {Mardon}}, \bibinfo
  {author} {\bibfnamefont {S.}~\bibnamefont {Rajendran}},\ and\ \bibinfo
  {author} {\bibfnamefont {W.~A.}\ \bibnamefont {Terrano}},\ }\href
  {https://doi.org/10.1103/PhysRevD.93.075029} {\bibfield  {journal} {\bibinfo
  {journal} {Phys. Rev. D}\ }\textbf {\bibinfo {volume} {93}},\ \bibinfo
  {pages} {075029} (\bibinfo {year} {2016})}\BibitemShut {NoStop}%
\bibitem [{\citenamefont {Badurina}\ \emph {et~al.}(2022)\citenamefont
  {Badurina}, \citenamefont {Blas},\ and\ \citenamefont
  {McCabe}}]{badurina2022refined}%
  \BibitemOpen
  \bibfield  {author} {\bibinfo {author} {\bibfnamefont {L.}~\bibnamefont
  {Badurina}}, \bibinfo {author} {\bibfnamefont {D.}~\bibnamefont {Blas}},\
  and\ \bibinfo {author} {\bibfnamefont {C.}~\bibnamefont {McCabe}},\ }\href
  {https://doi.org/10.1103/PHYSREVD.105.023006} {\bibfield  {journal} {\bibinfo
   {journal} {Phys. Rev. D}\ }\textbf {\bibinfo {volume} {105}},\ \bibinfo
  {pages} {023006} (\bibinfo {year} {2022})}\BibitemShut {NoStop}%
\bibitem [{\citenamefont {Arvanitaki}\ \emph {et~al.}(2018)\citenamefont
  {Arvanitaki}, \citenamefont {Graham}, \citenamefont {Hogan}, \citenamefont
  {Rajendran},\ and\ \citenamefont {{Van Tilburg}}}]{arvanitaki2018search}%
  \BibitemOpen
  \bibfield  {author} {\bibinfo {author} {\bibfnamefont {A.}~\bibnamefont
  {Arvanitaki}}, \bibinfo {author} {\bibfnamefont {P.~W.}\ \bibnamefont
  {Graham}}, \bibinfo {author} {\bibfnamefont {J.~M.}\ \bibnamefont {Hogan}},
  \bibinfo {author} {\bibfnamefont {S.}~\bibnamefont {Rajendran}},\ and\
  \bibinfo {author} {\bibfnamefont {K.}~\bibnamefont {{Van Tilburg}}},\ }\href
  {https://doi.org/10.1103/PHYSREVD.97.075020} {\bibfield  {journal} {\bibinfo
  {journal} {Phys. Rev. D}\ }\textbf {\bibinfo {volume} {97}},\ \bibinfo
  {pages} {075020} (\bibinfo {year} {2018})}\BibitemShut {NoStop}%
\bibitem [{\citenamefont {Pa{\v{s}}teka}\ \emph {et~al.}(2019)\citenamefont
  {Pa{\v{s}}teka}, \citenamefont {Hao}, \citenamefont {Borschevsky},
  \citenamefont {Flambaum},\ and\ \citenamefont
  {Schwerdtfeger}}]{pasteka2019material}%
  \BibitemOpen
  \bibfield  {author} {\bibinfo {author} {\bibfnamefont {L.~F.}\ \bibnamefont
  {Pa{\v{s}}teka}}, \bibinfo {author} {\bibfnamefont {Y.}~\bibnamefont {Hao}},
  \bibinfo {author} {\bibfnamefont {A.}~\bibnamefont {Borschevsky}}, \bibinfo
  {author} {\bibfnamefont {V.~V.}\ \bibnamefont {Flambaum}},\ and\ \bibinfo
  {author} {\bibfnamefont {P.}~\bibnamefont {Schwerdtfeger}},\ }\href
  {https://doi.org/10.1103/PHYSREVLETT.122.160801} {\bibfield  {journal}
  {\bibinfo  {journal} {Phys. Rev. Lett.}\ }\textbf {\bibinfo {volume} {122}},\
  \bibinfo {pages} {160801} (\bibinfo {year} {2019})}\BibitemShut {NoStop}%
\bibitem [{\citenamefont {Branca}\ \emph {et~al.}(2017)\citenamefont {Branca},
  \citenamefont {Bonaldi}, \citenamefont {Cerdonio}, \citenamefont {Conti},
  \citenamefont {Falferi}, \citenamefont {Marin}, \citenamefont {Mezzena},
  \citenamefont {Ortolan}, \citenamefont {Prodi}, \citenamefont {Taffarello},
  \citenamefont {Vedovato}, \citenamefont {Vinante}, \citenamefont {Vitale},\
  and\ \citenamefont {Zendri}}]{branca2017search}%
  \BibitemOpen
  \bibfield  {author} {\bibinfo {author} {\bibfnamefont {A.}~\bibnamefont
  {Branca}}, \bibinfo {author} {\bibfnamefont {M.}~\bibnamefont {Bonaldi}},
  \bibinfo {author} {\bibfnamefont {M.}~\bibnamefont {Cerdonio}}, \bibinfo
  {author} {\bibfnamefont {L.}~\bibnamefont {Conti}}, \bibinfo {author}
  {\bibfnamefont {P.}~\bibnamefont {Falferi}}, \bibinfo {author} {\bibfnamefont
  {F.}~\bibnamefont {Marin}}, \bibinfo {author} {\bibfnamefont
  {R.}~\bibnamefont {Mezzena}}, \bibinfo {author} {\bibfnamefont
  {A.}~\bibnamefont {Ortolan}}, \bibinfo {author} {\bibfnamefont {G.~A.}\
  \bibnamefont {Prodi}}, \bibinfo {author} {\bibfnamefont {L.}~\bibnamefont
  {Taffarello}}, \bibinfo {author} {\bibfnamefont {G.}~\bibnamefont
  {Vedovato}}, \bibinfo {author} {\bibfnamefont {A.}~\bibnamefont {Vinante}},
  \bibinfo {author} {\bibfnamefont {S.}~\bibnamefont {Vitale}},\ and\ \bibinfo
  {author} {\bibfnamefont {J.-P.}\ \bibnamefont {Zendri}},\ }\href
  {https://doi.org/10.1103/PhysRevLett.118.021302} {\bibfield  {journal}
  {\bibinfo  {journal} {Phys. Rev. Lett.}\ }\textbf {\bibinfo {volume} {118}},\
  \bibinfo {pages} {021302} (\bibinfo {year} {2017})}\BibitemShut {NoStop}%
\bibitem [{\citenamefont {Manley}\ \emph {et~al.}(2020)\citenamefont {Manley},
  \citenamefont {Wilson}, \citenamefont {Stump}, \citenamefont {Grin},\ and\
  \citenamefont {Singh}}]{manley2020searching}%
  \BibitemOpen
  \bibfield  {author} {\bibinfo {author} {\bibfnamefont {J.}~\bibnamefont
  {Manley}}, \bibinfo {author} {\bibfnamefont {D.~J.}\ \bibnamefont {Wilson}},
  \bibinfo {author} {\bibfnamefont {R.}~\bibnamefont {Stump}}, \bibinfo
  {author} {\bibfnamefont {D.}~\bibnamefont {Grin}},\ and\ \bibinfo {author}
  {\bibfnamefont {S.}~\bibnamefont {Singh}},\ }\href
  {https://doi.org/10.1103/PHYSREVLETT.124.151301} {\bibfield  {journal}
  {\bibinfo  {journal} {Phys. Rev. Lett.}\ }\textbf {\bibinfo {volume} {124}},\
  \bibinfo {pages} {151301} (\bibinfo {year} {2020})}\BibitemShut {NoStop}%
\bibitem [{\citenamefont {Stadnik}\ and\ \citenamefont
  {Flambaum}(2015{\natexlab{b}})}]{stadnik2015searching}%
  \BibitemOpen
  \bibfield  {author} {\bibinfo {author} {\bibfnamefont {Y.~V.}\ \bibnamefont
  {Stadnik}}\ and\ \bibinfo {author} {\bibfnamefont {V.~V.}\ \bibnamefont
  {Flambaum}},\ }\href {https://doi.org/10.1103/PHYSREVLETT.114.161301}
  {\bibfield  {journal} {\bibinfo  {journal} {Phys. Rev. Lett.}\ }\textbf
  {\bibinfo {volume} {114}},\ \bibinfo {pages} {161301} (\bibinfo {year}
  {2015}{\natexlab{b}})}\BibitemShut {NoStop}%
\bibitem [{\citenamefont {Geraci}\ \emph {et~al.}(2019)\citenamefont {Geraci},
  \citenamefont {Bradley}, \citenamefont {Gao}, \citenamefont {Weinstein},\
  and\ \citenamefont {Derevianko}}]{geraci2019searching}%
  \BibitemOpen
  \bibfield  {author} {\bibinfo {author} {\bibfnamefont {A.~A.}\ \bibnamefont
  {Geraci}}, \bibinfo {author} {\bibfnamefont {C.}~\bibnamefont {Bradley}},
  \bibinfo {author} {\bibfnamefont {D.}~\bibnamefont {Gao}}, \bibinfo {author}
  {\bibfnamefont {J.}~\bibnamefont {Weinstein}},\ and\ \bibinfo {author}
  {\bibfnamefont {A.}~\bibnamefont {Derevianko}},\ }\href
  {https://doi.org/10.1103/PhysRevLett.123.031304} {\bibfield  {journal}
  {\bibinfo  {journal} {Phys. Rev. Lett}\ }\textbf {\bibinfo {volume} {123}},\
  \bibinfo {pages} {031304} (\bibinfo {year} {2019})}\BibitemShut {NoStop}%
\bibitem [{\citenamefont {Kennedy}\ \emph {et~al.}(2020)\citenamefont
  {Kennedy}, \citenamefont {Oelker}, \citenamefont {Robinson}, \citenamefont
  {Bothwell}, \citenamefont {Kedar}, \citenamefont {Milner}, \citenamefont
  {Marti}, \citenamefont {Derevianko},\ and\ \citenamefont
  {Ye}}]{kennedy2020precision}%
  \BibitemOpen
  \bibfield  {author} {\bibinfo {author} {\bibfnamefont {C.~J.}\ \bibnamefont
  {Kennedy}}, \bibinfo {author} {\bibfnamefont {E.}~\bibnamefont {Oelker}},
  \bibinfo {author} {\bibfnamefont {J.~M.}\ \bibnamefont {Robinson}}, \bibinfo
  {author} {\bibfnamefont {T.}~\bibnamefont {Bothwell}}, \bibinfo {author}
  {\bibfnamefont {D.}~\bibnamefont {Kedar}}, \bibinfo {author} {\bibfnamefont
  {W.~R.}\ \bibnamefont {Milner}}, \bibinfo {author} {\bibfnamefont {G.~E.}\
  \bibnamefont {Marti}}, \bibinfo {author} {\bibfnamefont {A.}~\bibnamefont
  {Derevianko}},\ and\ \bibinfo {author} {\bibfnamefont {J.}~\bibnamefont
  {Ye}},\ }\href
  {https://journals.aps.org/prl/abstract/10.1103/PhysRevLett.125.201302}
  {\bibfield  {journal} {\bibinfo  {journal} {Phys. Rev. Lett.}\ }\textbf
  {\bibinfo {volume} {125}},\ \bibinfo {pages} {201302} (\bibinfo {year}
  {2020})}\BibitemShut {NoStop}%
\bibitem [{\citenamefont {Campbell}\ \emph {et~al.}(2021)\citenamefont
  {Campbell}, \citenamefont {Mcallister}, \citenamefont {Goryachev},
  \citenamefont {Ivanov},\ and\ \citenamefont {Tobar}}]{campbell2021searching}%
  \BibitemOpen
  \bibfield  {author} {\bibinfo {author} {\bibfnamefont {W.~M.}\ \bibnamefont
  {Campbell}}, \bibinfo {author} {\bibfnamefont {B.~T.}\ \bibnamefont
  {Mcallister}}, \bibinfo {author} {\bibfnamefont {M.}~\bibnamefont
  {Goryachev}}, \bibinfo {author} {\bibfnamefont {E.~N.}\ \bibnamefont
  {Ivanov}},\ and\ \bibinfo {author} {\bibfnamefont {M.~E.}\ \bibnamefont
  {Tobar}},\ }\href {https://doi.org/10.1103/PhysRevLett.126.071301} {\bibfield
   {journal} {\bibinfo  {journal} {Phys. Rev. Lett.}\ }\textbf {\bibinfo
  {volume} {126}},\ \bibinfo {pages} {71301} (\bibinfo {year}
  {2021})}\BibitemShut {NoStop}%
\bibitem [{\citenamefont {Savalle}\ \emph {et~al.}(2021)\citenamefont
  {Savalle}, \citenamefont {Hees}, \citenamefont {Frank}, \citenamefont
  {Cantin}, \citenamefont {Pottie}, \citenamefont {Roberts}, \citenamefont
  {Cros}, \citenamefont {McAllister},\ and\ \citenamefont
  {Wolf}}]{savalle2021searching}%
  \BibitemOpen
  \bibfield  {author} {\bibinfo {author} {\bibfnamefont {E.}~\bibnamefont
  {Savalle}}, \bibinfo {author} {\bibfnamefont {A.}~\bibnamefont {Hees}},
  \bibinfo {author} {\bibfnamefont {F.}~\bibnamefont {Frank}}, \bibinfo
  {author} {\bibfnamefont {E.}~\bibnamefont {Cantin}}, \bibinfo {author}
  {\bibfnamefont {P.~E.}\ \bibnamefont {Pottie}}, \bibinfo {author}
  {\bibfnamefont {B.~M.}\ \bibnamefont {Roberts}}, \bibinfo {author}
  {\bibfnamefont {L.}~\bibnamefont {Cros}}, \bibinfo {author} {\bibfnamefont
  {B.~T.}\ \bibnamefont {McAllister}},\ and\ \bibinfo {author} {\bibfnamefont
  {P.}~\bibnamefont {Wolf}},\ }\href
  {https://doi.org/10.1103/PHYSREVLETT.126.051301} {\bibfield  {journal}
  {\bibinfo  {journal} {Phys. Rev. Lett.}\ }\textbf {\bibinfo {volume} {126}},\
  \bibinfo {pages} {051301} (\bibinfo {year} {2021})}\BibitemShut {NoStop}%
\bibitem [{\citenamefont {Berlin}(2016)}]{berlin2016neutrino}%
  \BibitemOpen
  \bibfield  {author} {\bibinfo {author} {\bibfnamefont {A.}~\bibnamefont
  {Berlin}},\ }\href {https://doi.org/10.1103/PHYSREVLETT.117.231801}
  {\bibfield  {journal} {\bibinfo  {journal} {Phys. Rev. Lett.}\ }\textbf
  {\bibinfo {volume} {117}},\ \bibinfo {pages} {231801} (\bibinfo {year}
  {2016})}\BibitemShut {NoStop}%
\bibitem [{\citenamefont {Janish}\ and\ \citenamefont
  {Ramani}(2020)}]{janish2020muon}%
  \BibitemOpen
  \bibfield  {author} {\bibinfo {author} {\bibfnamefont {R.}~\bibnamefont
  {Janish}}\ and\ \bibinfo {author} {\bibfnamefont {H.}~\bibnamefont
  {Ramani}},\ }\href {https://doi.org/10.1103/PhysRevD.102.115018} {\bibfield
  {journal} {\bibinfo  {journal} {Phys. Rev. D}\ }\textbf {\bibinfo {volume}
  {102}},\ \bibinfo {pages} {115018} (\bibinfo {year} {2020})}\BibitemShut
  {NoStop}%
\bibitem [{\citenamefont {Vermeulen}\ \emph {et~al.}(2021)\citenamefont
  {Vermeulen}, \citenamefont {Relton}, \citenamefont {Grote}, \citenamefont
  {Raymond}, \citenamefont {Affeldt}, \citenamefont {Bergamin}, \citenamefont
  {Bisht}, \citenamefont {Brinkmann}, \citenamefont {Danzmann}, \citenamefont
  {Doravari}, \citenamefont {Kringel}, \citenamefont {Lough}, \citenamefont
  {L{\"{u}}ck}, \citenamefont {Mehmet}, \citenamefont {Mukund}, \citenamefont
  {Nadji}, \citenamefont {Schreiber}, \citenamefont {Sorazu}, \citenamefont
  {Strain}, \citenamefont {Vahlbruch}, \citenamefont {Weinert}, \citenamefont
  {Willke},\ and\ \citenamefont {Wittel}}]{vermeulen2021direct}%
  \BibitemOpen
  \bibfield  {author} {\bibinfo {author} {\bibfnamefont {S.~M.}\ \bibnamefont
  {Vermeulen}}, \bibinfo {author} {\bibfnamefont {P.}~\bibnamefont {Relton}},
  \bibinfo {author} {\bibfnamefont {H.}~\bibnamefont {Grote}}, \bibinfo
  {author} {\bibfnamefont {V.}~\bibnamefont {Raymond}}, \bibinfo {author}
  {\bibfnamefont {C.}~\bibnamefont {Affeldt}}, \bibinfo {author} {\bibfnamefont
  {F.}~\bibnamefont {Bergamin}}, \bibinfo {author} {\bibfnamefont
  {A.}~\bibnamefont {Bisht}}, \bibinfo {author} {\bibfnamefont
  {M.}~\bibnamefont {Brinkmann}}, \bibinfo {author} {\bibfnamefont
  {K.}~\bibnamefont {Danzmann}}, \bibinfo {author} {\bibfnamefont
  {S.}~\bibnamefont {Doravari}}, \bibinfo {author} {\bibfnamefont
  {V.}~\bibnamefont {Kringel}}, \bibinfo {author} {\bibfnamefont
  {J.}~\bibnamefont {Lough}}, \bibinfo {author} {\bibfnamefont
  {H.}~\bibnamefont {L{\"{u}}ck}}, \bibinfo {author} {\bibfnamefont
  {M.}~\bibnamefont {Mehmet}}, \bibinfo {author} {\bibfnamefont
  {N.}~\bibnamefont {Mukund}}, \bibinfo {author} {\bibfnamefont
  {S.}~\bibnamefont {Nadji}}, \bibinfo {author} {\bibfnamefont
  {E.}~\bibnamefont {Schreiber}}, \bibinfo {author} {\bibfnamefont
  {B.}~\bibnamefont {Sorazu}}, \bibinfo {author} {\bibfnamefont {K.~A.}\
  \bibnamefont {Strain}}, \bibinfo {author} {\bibfnamefont {H.}~\bibnamefont
  {Vahlbruch}}, \bibinfo {author} {\bibfnamefont {M.}~\bibnamefont {Weinert}},
  \bibinfo {author} {\bibfnamefont {B.}~\bibnamefont {Willke}},\ and\ \bibinfo
  {author} {\bibfnamefont {H.}~\bibnamefont {Wittel}},\ }\href
  {https://doi.org/10.1038/s41586-021-04031-y} {\bibfield  {journal} {\bibinfo
  {journal} {Nature}\ }\textbf {\bibinfo {volume} {600}},\ \bibinfo {pages}
  {424} (\bibinfo {year} {2021})}\BibitemShut {NoStop}%
\bibitem [{\citenamefont {Hees}\ \emph {et~al.}(2020)\citenamefont {Hees},
  \citenamefont {Do}, \citenamefont {Roberts}, \citenamefont {Ghez},
  \citenamefont {Nishiyama}, \citenamefont {Bentley}, \citenamefont {Gautam},
  \citenamefont {Jia}, \citenamefont {Kara}, \citenamefont {Lu}, \citenamefont
  {Saida}, \citenamefont {Sakai}, \citenamefont {Takahashi},\ and\
  \citenamefont {Takamori}}]{hees2020search}%
  \BibitemOpen
  \bibfield  {author} {\bibinfo {author} {\bibfnamefont {A.}~\bibnamefont
  {Hees}}, \bibinfo {author} {\bibfnamefont {T.}~\bibnamefont {Do}}, \bibinfo
  {author} {\bibfnamefont {B.~M.}\ \bibnamefont {Roberts}}, \bibinfo {author}
  {\bibfnamefont {A.~M.}\ \bibnamefont {Ghez}}, \bibinfo {author}
  {\bibfnamefont {S.}~\bibnamefont {Nishiyama}}, \bibinfo {author}
  {\bibfnamefont {R.~O.}\ \bibnamefont {Bentley}}, \bibinfo {author}
  {\bibfnamefont {A.~K.}\ \bibnamefont {Gautam}}, \bibinfo {author}
  {\bibfnamefont {S.}~\bibnamefont {Jia}}, \bibinfo {author} {\bibfnamefont
  {T.}~\bibnamefont {Kara}}, \bibinfo {author} {\bibfnamefont {J.~R.}\
  \bibnamefont {Lu}}, \bibinfo {author} {\bibfnamefont {H.}~\bibnamefont
  {Saida}}, \bibinfo {author} {\bibfnamefont {S.}~\bibnamefont {Sakai}},
  \bibinfo {author} {\bibfnamefont {M.}~\bibnamefont {Takahashi}},\ and\
  \bibinfo {author} {\bibfnamefont {Y.}~\bibnamefont {Takamori}},\ }\href
  {https://doi.org/10.1103/PhysRevLett.124.081101} {\bibfield  {journal}
  {\bibinfo  {journal} {Phys. Rev. Lett.}\ }\textbf {\bibinfo {volume} {124}},\
  \bibinfo {pages} {081101} (\bibinfo {year} {2020})}\BibitemShut {NoStop}%
\bibitem [{\citenamefont {Choi}\ and\ \citenamefont
  {Jung}(2019)}]{choi2019new}%
  \BibitemOpen
  \bibfield  {author} {\bibinfo {author} {\bibfnamefont {H.~G.}\ \bibnamefont
  {Choi}}\ and\ \bibinfo {author} {\bibfnamefont {S.}~\bibnamefont {Jung}},\
  }\href {https://doi.org/10.1103/PhysRevD.99.015013} {\bibfield  {journal}
  {\bibinfo  {journal} {Phys. Rev. D}\ }\textbf {\bibinfo {volume} {99}},\
  \bibinfo {pages} {015013} (\bibinfo {year} {2019})}\BibitemShut {NoStop}%
\bibitem [{\citenamefont {Kozyryev}\ \emph {et~al.}(2021)\citenamefont
  {Kozyryev}, \citenamefont {Lasner},\ and\ \citenamefont
  {Doyle}}]{kozyryev2021enhanced}%
  \BibitemOpen
  \bibfield  {author} {\bibinfo {author} {\bibfnamefont {I.}~\bibnamefont
  {Kozyryev}}, \bibinfo {author} {\bibfnamefont {Z.}~\bibnamefont {Lasner}},\
  and\ \bibinfo {author} {\bibfnamefont {J.~M.}\ \bibnamefont {Doyle}},\ }\href
  {https://doi.org/10.1103/PhysRevA.103.043313} {\bibfield  {journal} {\bibinfo
   {journal} {Phys. Rev. A}\ }\textbf {\bibinfo {volume} {103}},\ \bibinfo
  {pages} {043313} (\bibinfo {year} {2021})}\BibitemShut {NoStop}%
\bibitem [{\citenamefont {Hu}\ \emph {et~al.}(2000)\citenamefont {Hu},
  \citenamefont {Barkana},\ and\ \citenamefont {Gruzinov}}]{hu2000fuzzy}%
  \BibitemOpen
  \bibfield  {author} {\bibinfo {author} {\bibfnamefont {W.}~\bibnamefont
  {Hu}}, \bibinfo {author} {\bibfnamefont {R.}~\bibnamefont {Barkana}},\ and\
  \bibinfo {author} {\bibfnamefont {A.}~\bibnamefont {Gruzinov}},\ }\href
  {https://doi.org/10.1103/PhysRevLett.85.1158} {\bibfield  {journal} {\bibinfo
   {journal} {Phys. Rev. Lett.}\ }\textbf {\bibinfo {volume} {85}},\ \bibinfo
  {pages} {1158} (\bibinfo {year} {2000})}\BibitemShut {NoStop}%
\bibitem [{\citenamefont {Marsh}\ and\ \citenamefont
  {Pop}(2015)}]{marsh2015axion}%
  \BibitemOpen
  \bibfield  {author} {\bibinfo {author} {\bibfnamefont {D.~J.}\ \bibnamefont
  {Marsh}}\ and\ \bibinfo {author} {\bibfnamefont {A.-R.}\ \bibnamefont
  {Pop}},\ }\href {https://doi.org/10.1093/mnras/stv1050} {\bibfield  {journal}
  {\bibinfo  {journal} {Mon. Not. R. Astron. Soc.}\ }\textbf {\bibinfo {volume}
  {451}},\ \bibinfo {pages} {2479} (\bibinfo {year} {2015})}\BibitemShut
  {NoStop}%
\bibitem [{\citenamefont {Lora}\ \emph {et~al.}(2012)\citenamefont {Lora},
  \citenamefont {Magana}, \citenamefont {Bernal}, \citenamefont
  {S{\'a}nchez-Salcedo},\ and\ \citenamefont {Grebel}}]{lora2012mass}%
  \BibitemOpen
  \bibfield  {author} {\bibinfo {author} {\bibfnamefont {V.}~\bibnamefont
  {Lora}}, \bibinfo {author} {\bibfnamefont {J.}~\bibnamefont {Magana}},
  \bibinfo {author} {\bibfnamefont {A.}~\bibnamefont {Bernal}}, \bibinfo
  {author} {\bibfnamefont {F.}~\bibnamefont {S{\'a}nchez-Salcedo}},\ and\
  \bibinfo {author} {\bibfnamefont {E.}~\bibnamefont {Grebel}},\ }\href
  {https://doi.org/10.1088/1475-7516/2012/02/011} {\bibfield  {journal}
  {\bibinfo  {journal} {J. Cosmol. and Astropart. Phys.}\ }\textbf {\bibinfo
  {volume} {2012}},\ \bibinfo {pages} {011} (\bibinfo {year}
  {2012})}\BibitemShut {NoStop}%
\bibitem [{\citenamefont {Norrgard}\ \emph {et~al.}(2019)\citenamefont
  {Norrgard}, \citenamefont {Barker}, \citenamefont {Eckel}, \citenamefont
  {Fedchak}, \citenamefont {Klimov},\ and\ \citenamefont
  {Scherschligt}}]{norrgard2019nuclear}%
  \BibitemOpen
  \bibfield  {author} {\bibinfo {author} {\bibfnamefont {E.~B.}\ \bibnamefont
  {Norrgard}}, \bibinfo {author} {\bibfnamefont {D.~S.}\ \bibnamefont
  {Barker}}, \bibinfo {author} {\bibfnamefont {S.}~\bibnamefont {Eckel}},
  \bibinfo {author} {\bibfnamefont {J.~A.}\ \bibnamefont {Fedchak}}, \bibinfo
  {author} {\bibfnamefont {N.~N.}\ \bibnamefont {Klimov}},\ and\ \bibinfo
  {author} {\bibfnamefont {J.}~\bibnamefont {Scherschligt}},\ }\href
  {https://doi.org/10.1038/s42005-019-0181-1} {\bibfield  {journal} {\bibinfo
  {journal} {Commun. Phys.}\ }\textbf {\bibinfo {volume} {2}},\ \bibinfo
  {pages} {77} (\bibinfo {year} {2019})}\BibitemShut {NoStop}%
\bibitem [{\citenamefont {Kozyryev}\ and\ \citenamefont
  {Hutzler}(2017)}]{kozyryev2017precision}%
  \BibitemOpen
  \bibfield  {author} {\bibinfo {author} {\bibfnamefont {I.}~\bibnamefont
  {Kozyryev}}\ and\ \bibinfo {author} {\bibfnamefont {N.~R.}\ \bibnamefont
  {Hutzler}},\ }\href {https://doi.org/10.1103/PhysRevLett.119.133002}
  {\bibfield  {journal} {\bibinfo  {journal} {Phys. Rev. Lett.}\ }\textbf
  {\bibinfo {volume} {119}},\ \bibinfo {pages} {133002} (\bibinfo {year}
  {2017})}\BibitemShut {NoStop}%
\bibitem [{\citenamefont {Gaul}\ and\ \citenamefont
  {Berger}(2020)}]{gaul2020abinitio}%
  \BibitemOpen
  \bibfield  {author} {\bibinfo {author} {\bibfnamefont {K.}~\bibnamefont
  {Gaul}}\ and\ \bibinfo {author} {\bibfnamefont {R.}~\bibnamefont {Berger}},\
  }\href {https://doi.org/10.1103/PhysRevA.101.012508} {\bibfield  {journal}
  {\bibinfo  {journal} {Phys. Rev. A}\ }\textbf {\bibinfo {volume} {101}},\
  \bibinfo {pages} {12508} (\bibinfo {year} {2020})}\BibitemShut {NoStop}%
\bibitem [{\citenamefont {Kozyryev}\ \emph {et~al.}(2017)\citenamefont
  {Kozyryev}, \citenamefont {Baum}, \citenamefont {Matsuda}, \citenamefont
  {Augenbraun}, \citenamefont {Anderegg}, \citenamefont {Sedlack},\ and\
  \citenamefont {Doyle}}]{kozyryev2017sisyphus}%
  \BibitemOpen
  \bibfield  {author} {\bibinfo {author} {\bibfnamefont {I.}~\bibnamefont
  {Kozyryev}}, \bibinfo {author} {\bibfnamefont {L.}~\bibnamefont {Baum}},
  \bibinfo {author} {\bibfnamefont {K.}~\bibnamefont {Matsuda}}, \bibinfo
  {author} {\bibfnamefont {B.~L.}\ \bibnamefont {Augenbraun}}, \bibinfo
  {author} {\bibfnamefont {L.}~\bibnamefont {Anderegg}}, \bibinfo {author}
  {\bibfnamefont {A.~P.}\ \bibnamefont {Sedlack}},\ and\ \bibinfo {author}
  {\bibfnamefont {J.~M.}\ \bibnamefont {Doyle}},\ }\href
  {https://doi.org/10.1103/PhysRevLett.118.173201} {\bibfield  {journal}
  {\bibinfo  {journal} {Phys. Rev. Lett.}\ }\textbf {\bibinfo {volume} {118}},\
  \bibinfo {pages} {173201} (\bibinfo {year} {2017})}\BibitemShut {NoStop}%
\bibitem [{\citenamefont {Baum}\ \emph {et~al.}(2020)\citenamefont {Baum},
  \citenamefont {Vilas}, \citenamefont {Hallas}, \citenamefont {Augenbraun},
  \citenamefont {Raval}, \citenamefont {Mitra},\ and\ \citenamefont
  {Doyle}}]{Baum2020magneto}%
  \BibitemOpen
  \bibfield  {author} {\bibinfo {author} {\bibfnamefont {L.}~\bibnamefont
  {Baum}}, \bibinfo {author} {\bibfnamefont {N.~B.}\ \bibnamefont {Vilas}},
  \bibinfo {author} {\bibfnamefont {C.}~\bibnamefont {Hallas}}, \bibinfo
  {author} {\bibfnamefont {B.~L.}\ \bibnamefont {Augenbraun}}, \bibinfo
  {author} {\bibfnamefont {S.}~\bibnamefont {Raval}}, \bibinfo {author}
  {\bibfnamefont {D.}~\bibnamefont {Mitra}},\ and\ \bibinfo {author}
  {\bibfnamefont {J.~M.}\ \bibnamefont {Doyle}},\ }\href
  {https://doi.org/10.1103/PhysRevLett.124.133201} {\bibfield  {journal}
  {\bibinfo  {journal} {Phys. Rev. Lett.}\ }\textbf {\bibinfo {volume} {124}},\
  \bibinfo {pages} {133201} (\bibinfo {year} {2020})}\BibitemShut {NoStop}%
\bibitem [{\citenamefont {Augenbraun}\ \emph {et~al.}(2020)\citenamefont
  {Augenbraun}, \citenamefont {Lasner}, \citenamefont {Frenett}, \citenamefont
  {Sawaoka}, \citenamefont {Miller}, \citenamefont {Steimle},\ and\
  \citenamefont {Doyle}}]{augenbraun2020laser}%
  \BibitemOpen
  \bibfield  {author} {\bibinfo {author} {\bibfnamefont {B.~L.}\ \bibnamefont
  {Augenbraun}}, \bibinfo {author} {\bibfnamefont {Z.~D.}\ \bibnamefont
  {Lasner}}, \bibinfo {author} {\bibfnamefont {A.}~\bibnamefont {Frenett}},
  \bibinfo {author} {\bibfnamefont {H.}~\bibnamefont {Sawaoka}}, \bibinfo
  {author} {\bibfnamefont {C.}~\bibnamefont {Miller}}, \bibinfo {author}
  {\bibfnamefont {T.~C.}\ \bibnamefont {Steimle}},\ and\ \bibinfo {author}
  {\bibfnamefont {J.~M.}\ \bibnamefont {Doyle}},\ }\href
  {https://doi.org/10.1088/1367-2630/ab687b} {\bibfield  {journal} {\bibinfo
  {journal} {New J. Phys.}\ }\textbf {\bibinfo {volume} {22}},\ \bibinfo
  {pages} {022003} (\bibinfo {year} {2020})}\BibitemShut {NoStop}%
\bibitem [{\citenamefont {Mitra}\ \emph {et~al.}(2020)\citenamefont {Mitra},
  \citenamefont {Vilas}, \citenamefont {Hallas}, \citenamefont {Anderegg},
  \citenamefont {Augenbraun}, \citenamefont {Baum}, \citenamefont {Miller},
  \citenamefont {Raval},\ and\ \citenamefont {Doyle}}]{Mitra2020direct}%
  \BibitemOpen
  \bibfield  {author} {\bibinfo {author} {\bibfnamefont {D.}~\bibnamefont
  {Mitra}}, \bibinfo {author} {\bibfnamefont {N.~B.}\ \bibnamefont {Vilas}},
  \bibinfo {author} {\bibfnamefont {C.}~\bibnamefont {Hallas}}, \bibinfo
  {author} {\bibfnamefont {L.}~\bibnamefont {Anderegg}}, \bibinfo {author}
  {\bibfnamefont {B.~L.}\ \bibnamefont {Augenbraun}}, \bibinfo {author}
  {\bibfnamefont {L.}~\bibnamefont {Baum}}, \bibinfo {author} {\bibfnamefont
  {C.}~\bibnamefont {Miller}}, \bibinfo {author} {\bibfnamefont
  {S.}~\bibnamefont {Raval}},\ and\ \bibinfo {author} {\bibfnamefont {J.~M.}\
  \bibnamefont {Doyle}},\ }\href {https://www.doi.org/10.1126/science.abc5357}
  {\bibfield  {journal} {\bibinfo  {journal} {Science}\ }\textbf {\bibinfo
  {volume} {369}},\ \bibinfo {pages} {1366} (\bibinfo {year}
  {2020})}\BibitemShut {NoStop}%
\bibitem [{\citenamefont {Vilas}\ \emph {et~al.}(2021)\citenamefont {Vilas},
  \citenamefont {Hallas}, \citenamefont {Anderegg}, \citenamefont {Robichaud},
  \citenamefont {Winnicki}, \citenamefont {Mitra},\ and\ \citenamefont
  {Doyle}}]{vilas2021magneto}%
  \BibitemOpen
  \bibfield  {author} {\bibinfo {author} {\bibfnamefont {N.~B.}\ \bibnamefont
  {Vilas}}, \bibinfo {author} {\bibfnamefont {C.}~\bibnamefont {Hallas}},
  \bibinfo {author} {\bibfnamefont {L.}~\bibnamefont {Anderegg}}, \bibinfo
  {author} {\bibfnamefont {P.}~\bibnamefont {Robichaud}}, \bibinfo {author}
  {\bibfnamefont {A.}~\bibnamefont {Winnicki}}, \bibinfo {author}
  {\bibfnamefont {D.}~\bibnamefont {Mitra}},\ and\ \bibinfo {author}
  {\bibfnamefont {J.~M.}\ \bibnamefont {Doyle}},\ }\href
  {http://arxiv.org/abs/2112.08349} {\  (\bibinfo {year} {2021})},\ \Eprint
  {https://arxiv.org/abs/2112.08349} {arXiv:2112.08349} \BibitemShut {NoStop}%
\bibitem [{\citenamefont {Baum}\ \emph {et~al.}(2021)\citenamefont {Baum},
  \citenamefont {Vilas}, \citenamefont {Hallas}, \citenamefont {Augenbraun},
  \citenamefont {Raval}, \citenamefont {Mitra},\ and\ \citenamefont
  {Doyle}}]{Baum2020establishing}%
  \BibitemOpen
  \bibfield  {author} {\bibinfo {author} {\bibfnamefont {L.}~\bibnamefont
  {Baum}}, \bibinfo {author} {\bibfnamefont {N.~B.}\ \bibnamefont {Vilas}},
  \bibinfo {author} {\bibfnamefont {C.}~\bibnamefont {Hallas}}, \bibinfo
  {author} {\bibfnamefont {B.~L.}\ \bibnamefont {Augenbraun}}, \bibinfo
  {author} {\bibfnamefont {S.}~\bibnamefont {Raval}}, \bibinfo {author}
  {\bibfnamefont {D.}~\bibnamefont {Mitra}},\ and\ \bibinfo {author}
  {\bibfnamefont {J.~M.}\ \bibnamefont {Doyle}},\ }\href
  {https://doi.org/10.1103/PhysRevA.103.043111} {\bibfield  {journal} {\bibinfo
   {journal} {Phys. Rev. A}\ }\textbf {\bibinfo {volume} {103}},\ \bibinfo
  {pages} {043111} (\bibinfo {year} {2021})}\BibitemShut {NoStop}%
\bibitem [{\citenamefont {Zhang}\ \emph {et~al.}(2021)\citenamefont {Zhang},
  \citenamefont {Augenbraun}, \citenamefont {Lasner}, \citenamefont {Vilas},
  \citenamefont {Doyle},\ and\ \citenamefont {Cheng}}]{zhang2021accurate}%
  \BibitemOpen
  \bibfield  {author} {\bibinfo {author} {\bibfnamefont {C.}~\bibnamefont
  {Zhang}}, \bibinfo {author} {\bibfnamefont {B.}~\bibnamefont {Augenbraun}},
  \bibinfo {author} {\bibfnamefont {Z.~D.}\ \bibnamefont {Lasner}}, \bibinfo
  {author} {\bibfnamefont {N.~B.}\ \bibnamefont {Vilas}}, \bibinfo {author}
  {\bibfnamefont {J.~M.}\ \bibnamefont {Doyle}},\ and\ \bibinfo {author}
  {\bibfnamefont {L.}~\bibnamefont {Cheng}},\ }\href
  {https://doi.org/10.1063/5.0063611} {\bibfield  {journal} {\bibinfo
  {journal} {J. Chem. Phys.}\ }\textbf {\bibinfo {volume} {155}},\ \bibinfo
  {pages} {091101} (\bibinfo {year} {2021})}\BibitemShut {NoStop}%
\bibitem [{\citenamefont {K{\"o}ppel}\ \emph {et~al.}(1984)\citenamefont
  {K{\"o}ppel}, \citenamefont {Domcke},\ and\ \citenamefont
  {Cederbaum}}]{Koeppel84}%
  \BibitemOpen
  \bibfield  {author} {\bibinfo {author} {\bibfnamefont {H.}~\bibnamefont
  {K{\"o}ppel}}, \bibinfo {author} {\bibfnamefont {W.}~\bibnamefont {Domcke}},\
  and\ \bibinfo {author} {\bibfnamefont {L.~S.}\ \bibnamefont {Cederbaum}},\
  }\href {https://doi.org/10.1002/9780470142813.ch2} {\bibfield  {journal}
  {\bibinfo  {journal} {Adv. Chem. Phys.}\ }\textbf {\bibinfo {volume} {57}},\
  \bibinfo {pages} {59} (\bibinfo {year} {1984})}\BibitemShut {NoStop}%
\bibitem [{\citenamefont {Colbert}\ and\ \citenamefont
  {Miller}(1992)}]{Colbert92}%
  \BibitemOpen
  \bibfield  {author} {\bibinfo {author} {\bibfnamefont {D.~T.}\ \bibnamefont
  {Colbert}}\ and\ \bibinfo {author} {\bibfnamefont {W.~H.}\ \bibnamefont
  {Miller}},\ }\href {https://doi.org/10.1063/1.462100} {\bibfield  {journal}
  {\bibinfo  {journal} {J. Chem. Phys.}\ }\textbf {\bibinfo {volume} {96}},\
  \bibinfo {pages} {1982} (\bibinfo {year} {1992})}\BibitemShut {NoStop}%
\bibitem [{\citenamefont {Zhang}\ and\ \citenamefont
  {Cheng}(2020)}]{zhang2020performance}%
  \BibitemOpen
  \bibfield  {author} {\bibinfo {author} {\bibfnamefont {C.}~\bibnamefont
  {Zhang}}\ and\ \bibinfo {author} {\bibfnamefont {L.}~\bibnamefont {Cheng}},\
  }\href {https://doi.org/10.1080/00268976.2020.1768313} {\bibfield  {journal}
  {\bibinfo  {journal} {Mol. Phys.}\ }\textbf {\bibinfo {volume} {118}},\
  \bibinfo {pages} {e1768313} (\bibinfo {year} {2020})}\BibitemShut {NoStop}%
\bibitem [{\citenamefont {Ichino}\ \emph {et~al.}(2009)\citenamefont {Ichino},
  \citenamefont {Gauss},\ and\ \citenamefont {Stanton}}]{Ichino09}%
  \BibitemOpen
  \bibfield  {author} {\bibinfo {author} {\bibfnamefont {T.}~\bibnamefont
  {Ichino}}, \bibinfo {author} {\bibfnamefont {J.}~\bibnamefont {Gauss}},\ and\
  \bibinfo {author} {\bibfnamefont {J.~F.}\ \bibnamefont {Stanton}},\ }\href
  {https://doi.org/10.1063/1.3127246} {\bibfield  {journal} {\bibinfo
  {journal} {J. Chem. Phys.}\ }\textbf {\bibinfo {volume} {130}},\ \bibinfo
  {pages} {174105} (\bibinfo {year} {2009})}\BibitemShut {NoStop}%
\bibitem [{\citenamefont {Stanton}\ and\ \citenamefont
  {Bartlett}(1993)}]{Stanton93a}%
  \BibitemOpen
  \bibfield  {author} {\bibinfo {author} {\bibfnamefont {J.~F.}\ \bibnamefont
  {Stanton}}\ and\ \bibinfo {author} {\bibfnamefont {R.~J.}\ \bibnamefont
  {Bartlett}},\ }\href {https://doi.org/10.1063/1.464746} {\bibfield  {journal}
  {\bibinfo  {journal} {J. Chem. Phys.}\ }\textbf {\bibinfo {volume} {98}},\
  \bibinfo {pages} {7029} (\bibinfo {year} {1993})}\BibitemShut {NoStop}%
\bibitem [{\citenamefont {Nooijen}\ and\ \citenamefont
  {Bartlett}(1995)}]{Nooijen95}%
  \BibitemOpen
  \bibfield  {author} {\bibinfo {author} {\bibfnamefont {M.}~\bibnamefont
  {Nooijen}}\ and\ \bibinfo {author} {\bibfnamefont {R.~J.}\ \bibnamefont
  {Bartlett}},\ }\href {https://doi.org/10.1063/1.468592} {\bibfield  {journal}
  {\bibinfo  {journal} {J. Chem. Phys.}\ }\textbf {\bibinfo {volume} {102}},\
  \bibinfo {pages} {3629} (\bibinfo {year} {1995})}\BibitemShut {NoStop}%
\bibitem [{\citenamefont {Woon}\ and\ \citenamefont {{Dunning,
  Jr.}}(1995)}]{Woon95}%
  \BibitemOpen
  \bibfield  {author} {\bibinfo {author} {\bibfnamefont {D.~E.}\ \bibnamefont
  {Woon}}\ and\ \bibinfo {author} {\bibfnamefont {T.~H.}\ \bibnamefont
  {{Dunning, Jr.}}},\ }\href {https://doi.org/10.1063/1.470645} {\bibfield
  {journal} {\bibinfo  {journal} {J. Chem. Phys.}\ }\textbf {\bibinfo {volume}
  {103}},\ \bibinfo {pages} {4572} (\bibinfo {year} {1995})}\BibitemShut
  {NoStop}%
\bibitem [{\citenamefont {Hill}\ and\ \citenamefont
  {Peterson}(2017)}]{hill2017gaussian}%
  \BibitemOpen
  \bibfield  {author} {\bibinfo {author} {\bibfnamefont {J.~G.}\ \bibnamefont
  {Hill}}\ and\ \bibinfo {author} {\bibfnamefont {K.~A.}\ \bibnamefont
  {Peterson}},\ }\href {https://doi.org/10.1063/1.5010587} {\bibfield
  {journal} {\bibinfo  {journal} {J. Chem. Phys.}\ }\textbf {\bibinfo {volume}
  {147}},\ \bibinfo {pages} {244106} (\bibinfo {year} {2017})}\BibitemShut
  {NoStop}%
\bibitem [{\citenamefont {Dyall}(2001)}]{Dyall01}%
  \BibitemOpen
  \bibfield  {author} {\bibinfo {author} {\bibfnamefont {K.~G.}\ \bibnamefont
  {Dyall}},\ }\href {https://doi.org/10.1063/1.1413512} {\bibfield  {journal}
  {\bibinfo  {journal} {J. Chem. Phys.}\ }\textbf {\bibinfo {volume} {115}},\
  \bibinfo {pages} {9136} (\bibinfo {year} {2001})}\BibitemShut {NoStop}%
\bibitem [{\citenamefont {Liu}\ and\ \citenamefont {Peng}(2009)}]{Liu2009}%
  \BibitemOpen
  \bibfield  {author} {\bibinfo {author} {\bibfnamefont {W.}~\bibnamefont
  {Liu}}\ and\ \bibinfo {author} {\bibfnamefont {D.}~\bibnamefont {Peng}},\
  }\href {https://doi.org/10.1063/1.3159445} {\bibfield  {journal} {\bibinfo
  {journal} {J. Chem. Phys.}\ }\textbf {\bibinfo {volume} {131}},\ \bibinfo
  {pages} {031104} (\bibinfo {year} {2009})}\BibitemShut {NoStop}%
\bibitem [{\citenamefont {Cheng}\ and\ \citenamefont {Gauss}(2011)}]{Cheng11b}%
  \BibitemOpen
  \bibfield  {author} {\bibinfo {author} {\bibfnamefont {L.}~\bibnamefont
  {Cheng}}\ and\ \bibinfo {author} {\bibfnamefont {J.}~\bibnamefont {Gauss}},\
  }\href {https://doi.org/10.1063/1.3624397} {\bibfield  {journal} {\bibinfo
  {journal} {J. Chem. Phys.}\ }\textbf {\bibinfo {volume} {135}},\ \bibinfo
  {pages} {084114} (\bibinfo {year} {2011})}\BibitemShut {NoStop}%
\bibitem [{\citenamefont {Presunka}\ and\ \citenamefont
  {Coxon}(1995)}]{presunka1995laser}%
  \BibitemOpen
  \bibfield  {author} {\bibinfo {author} {\bibfnamefont {P.~I.}\ \bibnamefont
  {Presunka}}\ and\ \bibinfo {author} {\bibfnamefont {J.~A.}\ \bibnamefont
  {Coxon}},\ }\href {https://doi.org/10.1016/0301-0104(94)00330-D} {\bibfield
  {journal} {\bibinfo  {journal} {Chem. Phys.}\ }\textbf {\bibinfo {volume}
  {190}},\ \bibinfo {pages} {97} (\bibinfo {year} {1995})}\BibitemShut
  {NoStop}%
\bibitem [{\citenamefont {Nakagawa}\ \emph {et~al.}(1983)\citenamefont
  {Nakagawa}, \citenamefont {Wormsbecher},\ and\ \citenamefont
  {Harris}}]{nakagawa1983high}%
  \BibitemOpen
  \bibfield  {author} {\bibinfo {author} {\bibfnamefont {J.}~\bibnamefont
  {Nakagawa}}, \bibinfo {author} {\bibfnamefont {R.~F.}\ \bibnamefont
  {Wormsbecher}},\ and\ \bibinfo {author} {\bibfnamefont {D.~O.}\ \bibnamefont
  {Harris}},\ }\href {https://doi.org/10.1016/0022-2852(83)90336-3} {\bibfield
  {journal} {\bibinfo  {journal} {J. Mol. Spectrosc.}\ }\textbf {\bibinfo
  {volume} {97}},\ \bibinfo {pages} {37} (\bibinfo {year} {1983})}\BibitemShut
  {NoStop}%
\bibitem [{\citenamefont {Matthews}\ \emph {et~al.}(2020)\citenamefont
  {Matthews}, \citenamefont {Cheng}, \citenamefont {Harding}, \citenamefont
  {Lipparini}, \citenamefont {Stopkowicz}, \citenamefont {Jagau}, \citenamefont
  {Szalay}, \citenamefont {Gauss},\ and\ \citenamefont
  {Stanton}}]{Matthews2020a}%
  \BibitemOpen
  \bibfield  {author} {\bibinfo {author} {\bibfnamefont {D.~A.}\ \bibnamefont
  {Matthews}}, \bibinfo {author} {\bibfnamefont {L.}~\bibnamefont {Cheng}},
  \bibinfo {author} {\bibfnamefont {M.~E.}\ \bibnamefont {Harding}}, \bibinfo
  {author} {\bibfnamefont {F.}~\bibnamefont {Lipparini}}, \bibinfo {author}
  {\bibfnamefont {S.}~\bibnamefont {Stopkowicz}}, \bibinfo {author}
  {\bibfnamefont {T.-C.}\ \bibnamefont {Jagau}}, \bibinfo {author}
  {\bibfnamefont {P.~G.}\ \bibnamefont {Szalay}}, \bibinfo {author}
  {\bibfnamefont {J.}~\bibnamefont {Gauss}},\ and\ \bibinfo {author}
  {\bibfnamefont {J.~F.}\ \bibnamefont {Stanton}},\ }\href
  {https://doi.org/10.1063/5.0004837} {\bibfield  {journal} {\bibinfo
  {journal} {J. Chem. Phys.}\ }\textbf {\bibinfo {volume} {152}},\ \bibinfo
  {pages} {214108} (\bibinfo {year} {2020})}\BibitemShut {NoStop}%
\bibitem [{\citenamefont {Stanton}\ \emph {et~al.}()\citenamefont {Stanton},
  \citenamefont {Gauss}, \citenamefont {Cheng}, \citenamefont {Harding},
  \citenamefont {Matthews},\ and\ \citenamefont {Szalay}}]{cfour}%
  \BibitemOpen
  \bibfield  {author} {\bibinfo {author} {\bibfnamefont {J.~F.}\ \bibnamefont
  {Stanton}}, \bibinfo {author} {\bibfnamefont {J.}~\bibnamefont {Gauss}},
  \bibinfo {author} {\bibfnamefont {L.}~\bibnamefont {Cheng}}, \bibinfo
  {author} {\bibfnamefont {M.~E.}\ \bibnamefont {Harding}}, \bibinfo {author}
  {\bibfnamefont {D.~A.}\ \bibnamefont {Matthews}},\ and\ \bibinfo {author}
  {\bibfnamefont {P.~G.}\ \bibnamefont {Szalay}},\ }\href@noop {} {\bibinfo
  {title} {{CFOUR, Coupled-Cluster techniques for Computational Chemistry, a
  quantum-chemical program package}}},\ \bibinfo {note} {{W}ith contributions
  from {A}. {A}sthana, {A}.{A}. {A}uer, {R}.{J}. {B}artlett, {U}. {B}enedikt,
  {C}. {B}erger, {D}.{E}. {B}ernholdt, {S}. {B}laschke, {Y}. {J}. {B}omble,
  {S}. {B}urger, {O}. {C}hristiansen, {D}. {D}atta, {F}. {E}ngel, {R}. {F}aber,
  {J}. {G}reiner, {M}. {H}eckert, {O}. {H}eun, {M}. Hilgenberg, {C}. {H}uber,
  {T}.-{C}. {J}agau, {D}. {J}onsson, {J}. {J}us{\'e}lius, {T}. Kirsch,
  {M}.-{P}. {K}itsaras, {K}. {K}lein, {G}.{M}. {K}opper, {W}.{J}. {L}auderdale,
  {F}. {L}ipparini, {J}. {L}iu, {T}. {M}etzroth, {L}.{A}. {M}{\"u}ck, {D}.{P}.
  {O}'{N}eill, {T}. {N}ottoli, {J}. {O}swald, {D}.{R}. {P}rice, {E}.
  {P}rochnow, {C}. {P}uzzarini, {K}. {R}uud, {F}. {S}chiffmann, {W}.
  {S}chwalbach, {C}. {S}immons, {S}. {S}topkowicz, {A}. {T}ajti, {J}.
  {V}{\'a}zquez, {F}. {W}ang, {J}.{D}. {W}atts, {C}. {Z}hang, {X}. {Z}heng, and
  the integral packages {MOLECULE} ({J}. {A}lml{\"o}f and {P}.{R}. {T}aylor),
  {PROPS} ({P}.{R}. {T}aylor), {ABACUS} ({T}. {H}elgaker, {H}.{J}. {A}a.
  {J}ensen, {P}. {J}{\o}rgensen, and {J}. {O}lsen), and {ECP} routines by {A}.
  {V}. {M}itin and {C}. van {W}{\"u}llen. {F}or the current version, see
  \href{http://www.cfour.de}{http://www.cfour.de}.}\BibitemShut {Stop}%
\bibitem [{\citenamefont {Nguyen}\ \emph {et~al.}(2018)\citenamefont {Nguyen},
  \citenamefont {Steimle}, \citenamefont {Kozyryev}, \citenamefont {Huang},\
  and\ \citenamefont {McCoy}}]{nguyen2018fluorescence}%
  \BibitemOpen
  \bibfield  {author} {\bibinfo {author} {\bibfnamefont {D.-T.}\ \bibnamefont
  {Nguyen}}, \bibinfo {author} {\bibfnamefont {T.~C.}\ \bibnamefont {Steimle}},
  \bibinfo {author} {\bibfnamefont {I.}~\bibnamefont {Kozyryev}}, \bibinfo
  {author} {\bibfnamefont {M.}~\bibnamefont {Huang}},\ and\ \bibinfo {author}
  {\bibfnamefont {A.~B.}\ \bibnamefont {McCoy}},\ }\href
  {https://doi.org/10.1016/j.jms.2018.02.007} {\bibfield  {journal} {\bibinfo
  {journal} {J. Mol. Spectrosc.}\ }\textbf {\bibinfo {volume} {347}},\ \bibinfo
  {pages} {7} (\bibinfo {year} {2018})}\BibitemShut {NoStop}%
\bibitem [{\citenamefont {Presunka}\ and\ \citenamefont
  {Coxon}(1993)}]{presunka1993high}%
  \BibitemOpen
  \bibfield  {author} {\bibinfo {author} {\bibfnamefont {P.~I.}\ \bibnamefont
  {Presunka}}\ and\ \bibinfo {author} {\bibfnamefont {J.~A.}\ \bibnamefont
  {Coxon}},\ }\href {https://doi.org/10.1139/v93-211} {\bibfield  {journal}
  {\bibinfo  {journal} {Canadian Journal of Chemistry}\ }\textbf {\bibinfo
  {volume} {71}},\ \bibinfo {pages} {1689} (\bibinfo {year}
  {1993})}\BibitemShut {NoStop}%
\end{thebibliography}%
\end{document}